\documentclass[aps,prd,twocolumn,superscriptaddress,showpacs,preprintnumbers,amsmath,amssymb]{revtex4}

\usepackage{graphicx} 
\usepackage{dcolumn}  

\graphicspath{{ps}}

\usepackage{multirow,tabularx}

\begin{document}


\title{ \quad\\[1.0cm] Study of $e^+e^-\to \Upsilon(\rm1S,2S)\eta$ and $e^+e^-\to \Upsilon(\rm 1S)\eta^{\prime}$ at $\sqrt{s}=10.866$ GeV with the Belle detector}

\noaffiliation
\affiliation{Department of Physics, University of the Basque Country UPV/EHU, 48080 Bilbao}
\affiliation{University of Bonn, 53115 Bonn}
\affiliation{Brookhaven National Laboratory, Upton, New York 11973}
\affiliation{Budker Institute of Nuclear Physics SB RAS, Novosibirsk 630090}
\affiliation{Faculty of Mathematics and Physics, Charles University, 121 16 Prague}
\affiliation{Chonnam National University, Gwangju 61186}
\affiliation{University of Cincinnati, Cincinnati, Ohio 45221}
\affiliation{Deutsches Elektronen--Synchrotron, 22607 Hamburg}
\affiliation{Department of Physics, Fu Jen Catholic University, Taipei 24205}
\affiliation{Key Laboratory of Nuclear Physics and Ion-beam Application (MOE) and Institute of Modern Physics, Fudan University, Shanghai 200443}
\affiliation{Justus-Liebig-Universit\"at Gie\ss{}en, 35392 Gie\ss{}en}
\affiliation{Gifu University, Gifu 501-1193}
\affiliation{II. Physikalisches Institut, Georg-August-Universit\"at G\"ottingen, 37073 G\"ottingen}
\affiliation{SOKENDAI (The Graduate University for Advanced Studies), Hayama 240-0193}
\affiliation{Gyeongsang National University, Jinju 52828}
\affiliation{Department of Physics and Institute of Natural Sciences, Hanyang University, Seoul 04763}
\affiliation{University of Hawaii, Honolulu, Hawaii 96822}
\affiliation{High Energy Accelerator Research Organization (KEK), Tsukuba 305-0801}
\affiliation{J-PARC Branch, KEK Theory Center, High Energy Accelerator Research Organization (KEK), Tsukuba 305-0801}
\affiliation{National Research University Higher School of Economics, Moscow 101000}
\affiliation{Forschungszentrum J\"{u}lich, 52425 J\"{u}lich}
\affiliation{IKERBASQUE, Basque Foundation for Science, 48013 Bilbao}
\affiliation{Indian Institute of Science Education and Research Mohali, SAS Nagar, 140306}
\affiliation{Indian Institute of Technology Bhubaneswar, Satya Nagar 751007}
\affiliation{Indian Institute of Technology Hyderabad, Telangana 502285}
\affiliation{Indian Institute of Technology Madras, Chennai 600036}
\affiliation{Indiana University, Bloomington, Indiana 47408}
\affiliation{Institute of High Energy Physics, Chinese Academy of Sciences, Beijing 100049}
\affiliation{Institute of High Energy Physics, Vienna 1050}
\affiliation{Institute for High Energy Physics, Protvino 142281}
\affiliation{INFN - Sezione di Napoli, 80126 Napoli}
\affiliation{INFN - Sezione di Torino, 10125 Torino}
\affiliation{Advanced Science Research Center, Japan Atomic Energy Agency, Naka 319-1195}
\affiliation{J. Stefan Institute, 1000 Ljubljana}
\affiliation{Institut f\"ur Experimentelle Teilchenphysik, Karlsruher Institut f\"ur Technologie, 76131 Karlsruhe}
\affiliation{Kitasato University, Sagamihara 252-0373}
\affiliation{Korea Institute of Science and Technology Information, Daejeon 34141}
\affiliation{Korea University, Seoul 02841}
\affiliation{Kyoto Sangyo University, Kyoto 603-8555}
\affiliation{Kyungpook National University, Daegu 41566}
\affiliation{Universit\'{e} Paris-Saclay, CNRS/IN2P3, IJCLab, 91405 Orsay}
\affiliation{P.N. Lebedev Physical Institute of the Russian Academy of Sciences, Moscow 119991}
\affiliation{Faculty of Mathematics and Physics, University of Ljubljana, 1000 Ljubljana}
\affiliation{Ludwig Maximilians University, 80539 Munich}
\affiliation{Luther College, Decorah, Iowa 52101}
\affiliation{Malaviya National Institute of Technology Jaipur, Jaipur 302017}
\affiliation{Faculty of Chemistry and Chemical Engineering, University of Maribor, 2000 Maribor}
\affiliation{Max-Planck-Institut f\"ur Physik, 80805 M\"unchen}
\affiliation{School of Physics, University of Melbourne, Victoria 3010}
\affiliation{University of Mississippi, University, Mississippi 38677}
\affiliation{University of Miyazaki, Miyazaki 889-2192}
\affiliation{Moscow Physical Engineering Institute, Moscow 115409}
\affiliation{Graduate School of Science, Nagoya University, Nagoya 464-8602}
\affiliation{Kobayashi-Maskawa Institute, Nagoya University, Nagoya 464-8602}
\affiliation{Universit\`{a} di Napoli Federico II, 80126 Napoli}
\affiliation{Nara Women's University, Nara 630-8506}
\affiliation{National United University, Miao Li 36003}
\affiliation{Department of Physics, National Taiwan University, Taipei 10617}
\affiliation{H. Niewodniczanski Institute of Nuclear Physics, Krakow 31-342}
\affiliation{Nippon Dental University, Niigata 951-8580}
\affiliation{Niigata University, Niigata 950-2181}
\affiliation{Novosibirsk State University, Novosibirsk 630090}
\affiliation{Osaka City University, Osaka 558-8585}
\affiliation{Pacific Northwest National Laboratory, Richland, Washington 99352}
\affiliation{Panjab University, Chandigarh 160014}
\affiliation{Peking University, Beijing 100871}
\affiliation{University of Pittsburgh, Pittsburgh, Pennsylvania 15260}
\affiliation{Punjab Agricultural University, Ludhiana 141004}
\affiliation{Research Center for Nuclear Physics, Osaka University, Osaka 567-0047}
\affiliation{Meson Science Laboratory, Cluster for Pioneering Research, RIKEN, Saitama 351-0198}
\affiliation{Department of Modern Physics and State Key Laboratory of Particle Detection and Electronics, University of Science and Technology of China, Hefei 230026}
\affiliation{Seoul National University, Seoul 08826}
\affiliation{Showa Pharmaceutical University, Tokyo 194-8543}
\affiliation{Soongsil University, Seoul 06978}
\affiliation{Sungkyunkwan University, Suwon 16419}
\affiliation{School of Physics, University of Sydney, New South Wales 2006}
\affiliation{Department of Physics, Faculty of Science, University of Tabuk, Tabuk 71451}
\affiliation{Tata Institute of Fundamental Research, Mumbai 400005}
\affiliation{Department of Physics, Technische Universit\"at M\"unchen, 85748 Garching}
\affiliation{Toho University, Funabashi 274-8510}
\affiliation{Department of Physics, Tohoku University, Sendai 980-8578}
\affiliation{Earthquake Research Institute, University of Tokyo, Tokyo 113-0032}
\affiliation{Department of Physics, University of Tokyo, Tokyo 113-0033}
\affiliation{Tokyo Institute of Technology, Tokyo 152-8550}
\affiliation{Tokyo Metropolitan University, Tokyo 192-0397}
\affiliation{Virginia Polytechnic Institute and State University, Blacksburg, Virginia 24061}
\affiliation{Wayne State University, Detroit, Michigan 48202}
\affiliation{Yamagata University, Yamagata 990-8560}
\affiliation{Yonsei University, Seoul 03722}
\author{E.~Kovalenko}\affiliation{Budker Institute of Nuclear Physics SB RAS, Novosibirsk 630090}\affiliation{Novosibirsk State University, Novosibirsk 630090} 
\author{A.~Garmash}\affiliation{Budker Institute of Nuclear Physics SB RAS, Novosibirsk 630090}\affiliation{Novosibirsk State University, Novosibirsk 630090} 
\author{P.~Krokovny}\affiliation{Budker Institute of Nuclear Physics SB RAS, Novosibirsk 630090}\affiliation{Novosibirsk State University, Novosibirsk 630090} 
  \author{I.~Adachi}\affiliation{High Energy Accelerator Research Organization (KEK), Tsukuba 305-0801}\affiliation{SOKENDAI (The Graduate University for Advanced Studies), Hayama 240-0193} 
  \author{H.~Aihara}\affiliation{Department of Physics, University of Tokyo, Tokyo 113-0033} 
  \author{D.~M.~Asner}\affiliation{Brookhaven National Laboratory, Upton, New York 11973} 
  \author{V.~Aulchenko}\affiliation{Budker Institute of Nuclear Physics SB RAS, Novosibirsk 630090}\affiliation{Novosibirsk State University, Novosibirsk 630090} 
  \author{T.~Aushev}\affiliation{National Research University Higher School of Economics, Moscow 101000} 
  \author{R.~Ayad}\affiliation{Department of Physics, Faculty of Science, University of Tabuk, Tabuk 71451} 
  \author{V.~Babu}\affiliation{Deutsches Elektronen--Synchrotron, 22607 Hamburg} 
% ...
\collaboration{The Belle Collaboration}
     

\begin{abstract}
  We report the first observation of the processes $e^+e^-\to\Upsilon(\rm 1S,2S)\eta$ at 
  $\sqrt{s}=10.866$~GeV with a $10.2\sigma$ and $16.5\sigma$ significance respectively. 
  The measured Born cross sections are
  $\sigma(e^+e^- \to \Upsilon(2S)\eta)=2.07 \pm 0.21 \pm 0.19$~pb,
  and 
  $\sigma(e^+e^- \to \Upsilon(\rm 1S)\eta)=0.42 \pm 0.08 \pm 0.04$~pb.
We also set the upper limit on the cross section of the process
  $e^+e^- \to \Upsilon(\rm 1S)\eta^{\prime}$ to be
  $\sigma(e^+e^- \to \Upsilon(\rm 1S)\eta^{\prime})<0.035$~pb at $90\%$ CL.
  The results are obtained with the data sample collected 
with
  the Belle detector at the KEKB asymmetric-energy $e^+e^-$ collider
  in the energy range from $10.63$~GeV to $11.02$~GeV.
\end{abstract}

\pacs{14.40.Pq, 13.25.Gv, 13.66.Bc}

\maketitle


{\renewcommand{\thefootnote}{\fnsymbol{footnote}}}
\setcounter{footnote}{0}


\section{Introduction} \label{chapt0}

Bottomonium states (bound states of $b\Bar{b}$) above the $B\Bar{B}$ threshold have unexpected properties. 
For example, the $\Upsilon(\rm 10860)$ resonance, commonly denoted as $\Upsilon(\rm 5S)$, decays into $\Upsilon(\rm nS)\pi^+\pi^- ~(n=1,2,3)$ with widths around $300-400$~keV, about two orders of magnitude larger than those for similar decays of the $\Upsilon(\rm 2S)-\Upsilon(\rm 4S)$ which have widths around $0.5-5$ keV~\cite{PDG}.
One possible interpretation of such behavior is the existence of a light-flavor admixture in the $\Upsilon(\rm 5S)$ resonance~\cite{Voloshin:2011hw, Voloshin:2012dk}, which leads to cancellation of the suppression caused by heavy quark gluon emission. 

Observation by the Belle collaboration of unexpectedly large values for the ratios $\frac{\Gamma(\Upsilon(\rm 5S)\to h_b(1P)\pi^+\pi^-)}{\Gamma(\Upsilon(\rm 5S)\to\Upsilon(\rm 1S)\pi^+\pi^-)}=0.46\pm0.08^{+0.07}_{-0.12}$ and $\frac{\Gamma(\Upsilon(\rm 5S)\to h_b(2P)\pi^+\pi^-)}{\Gamma(\Upsilon(\rm 5S)\to\Upsilon(\rm 2S)\pi^+\pi^-)}=0.77\pm0.08^{+0.22}_{-0.17}$~\cite{Adachi:2011ji}, while it was expected to be ${\cal O}(10^{-2})$ due to heavy quark spin flip \cite{Voloshin:1980zf}, has led to discovery of exotic four-quark bound states $Z_b(10610)$ and $Z_b(10650)$~\cite{Bondar:2011aa}.
Another similar ratio $\frac{\Gamma(\Upsilon(\rm 4S,5S)\to \Upsilon(\rm 1S)\eta)}{\Gamma(\Upsilon(\rm 4S,5S)\to\Upsilon(\rm 1S)\pi^+\pi^-)}$ is also expected to be ${\cal O}(10^{-2})$ in the QCDME model \cite{Voloshin:2007dx}, but has been measured to be $2.41\pm0.40\pm0.12$ for the $\Upsilon(\rm 4S)$ resonance~\cite{Aubert:2008az}.
Moreover, the measurement of $\mathcal{B}(\Upsilon(\rm 4S)\to\eta h_b(1P))=(2.18\pm0.11\pm0.18)\times 10^{-3}$ \cite{Tamponi:2015xzb} violates naive quark-antiquark models \cite{Guo:2010ca} like QCDME.
Nevertheless, for bottomonium states below the $B\Bar{B}$ threshold, the QCDME model predictions are consistent with measurements: $\frac{\Gamma(\Upsilon(\rm 2S)\to \Upsilon(\rm 1S)\eta)}{\Gamma(\Upsilon(\rm 2S)\to\Upsilon(\rm 1S)\pi^+\pi^-)}=(1.64 \pm 0.25)\times 10^{-3}$~\cite{PDG} and $\frac{\Gamma(\Upsilon(\rm 3S)\to \Upsilon(\rm 1S)\eta)}{\Gamma(\Upsilon(\rm 3S)\to\Upsilon(\rm 1S)\pi^+\pi^-)}<2.3 \times 10^{-3}$~\cite{BABAR:2011ab}.
Therefore, analysis of similar processes is crucial for better understanding of the quark structure of bottomonium states above the $B\Bar{B}$ threshold.

This paper describes the study of hadronic transitions between bottomonium states with emission of an $\eta^{(\prime)}$ meson at $\sqrt{s}=10.866$ GeV.
The process $e^+e^-\to\Upsilon(\rm 2S)\eta$ is studied in two different modes: the first decay chain $\Upsilon(\rm 2S)\to\Upsilon(\rm 1S)\pi^+\pi^-$, $\Upsilon(\rm 1S)\to\mu^+\mu^-$, $\eta\to\gamma\gamma$ denoted as $\Upsilon(\rm 2S)\eta[\gamma\gamma]$; the second decay chain $\Upsilon(\rm 2S)\to\mu^+\mu^-$, $\eta\to\pi^+\pi^-\pi^0$, $\pi^0\to\gamma\gamma$ denoted as $\Upsilon(\rm 2S)\eta[3\pi]$.
The process $e^+e^-\to\Upsilon(\rm 1S)\eta$ is studied in the decay chain $\Upsilon(\rm 1S)\to\mu^+\mu^-$, $\eta\to\pi^+\pi^-\pi^0$, $\pi^0\to\gamma\gamma$ denoted as $\Upsilon(\rm 1S)\eta[3\pi]$. 
The process $e^+e^-\to\Upsilon(\rm 1S)\eta^{\prime}$ is studied in two different modes: the first decay chain $\Upsilon(\rm 1S)\to\mu^+\mu^-$, $\eta^{\prime}\to\pi^+\pi^-\eta$, $\eta\to\gamma\gamma$ denoted as $\Upsilon(\rm 1S)\eta^{\prime}[\pi\pi\eta]$; the second decay chain $\Upsilon(\rm 1S)\to\mu^+\mu^-$, $\eta^{\prime}\to\rho^0\gamma$, $\rho^0\to\pi^+\pi^-$ denoted as $\Upsilon(\rm 1S)\eta^{\prime}[\rho\gamma]$ and is the only process with the $\mu^+\mu^-\pi^+\pi^-\gamma$ final state, while other processes lead to the $\mu^+\mu^-\pi^+\pi^-\gamma\gamma$ final state.

A first evidence for the $e^+e^- \to \Upsilon(\rm 2S)\eta$ process has been reported in Ref. \cite{Tamponi:2018cuf}, where inclusive measurement with recoil mass distribution against $\eta$ meson was performed, the Born cross section (see eq. \ref{born crossection}) being $\sigma_{\rm B}(e^+e^- \to \Upsilon(\rm 2S)\eta)=1.02 \pm 0.30 \pm 0.17$~pb and the upper limit $\sigma_{\rm B}(e^+e^- \to \Upsilon(\rm 1S)\eta)<0.49$~pb being set at $90\%$ confidence level.
This analysis is exclusive in $\eta$ decays and independent from the latter one.

We use the data sample of $118.3$~fb$^{-1}$ collected at the $\Upsilon(\rm 5S)$ resonance and the data sample of $21$~fb$^{-1}$ collected during the energy scan in the range from $10.63$~GeV to $11.02$~GeV by the Belle detector~\cite{Abashian:2000cg, Brodzicka:2012jm} at the KEKB asymmetric-energy $e^+e^-$ collider \cite{Kurokawa:2001nw, Abe:2013kxa}.
The average center-of-mass (CM) energy of the $\Upsilon(\rm 5S)$ sample is $\sqrt{s} = 10.866$~GeV. 
The Belle detector was a large-solid-angle magnetic spectrometer that consisted of a silicon vertex detector, a 50-layer central drift chamber (CDC), an array of aerogel threshold Cherenkov counters (ACC), a barrel-like arrangement of time-of-flight scintillation counters, and an electromagnetic calorimeter comprised of CsI(Tl) crystals (ECL) located inside a superconducting solenoid coil that provided a 1.5 T magnetic field. 
An iron flux-return yoke located outside of the coil (KLM) was instrumented to detect $K_L^0$ mesons and to identify muons.

Event selection requirements are optimized using a full Monte Carlo (MC) simulation.
MC events are generated using EvtGen~\cite{Lange:2001uf} and the detector response is modeled using GEANT3~\cite{Brun:1987ma}.
In the simulation of $e^+e^-\to \Upsilon(\rm 1S,2S)\eta^{(\prime)}$ we use the angular distribution dictated by the quantum numbers for a vector decay to a pseudoscalar and a vector.
The dimuon decay of $\Upsilon(\rm 1S,2S)$ is simulated to be distributed uniformly in phase space, taking into account the proper spin dynamics for decay of a massive vector meson to two leptons.
For $\Upsilon(\rm 2S)\to \Upsilon(\rm 1S)\pi^+\pi^-$, we use a dipion invariant mass distribution according to the Voloshin and Zakharov model \cite{Voloshin:1980zf} measured in~\cite{Alexander:1998dq}.
For the $\eta\to\pi^+\pi^-\pi^0$ decay, final state particles are distributed in phase space according to the model from~\cite{Ambrosino:2008ht}.
Other decays are generated uniformly in phase space.
Final-state radiation is taken into account using the PHOTOS package~\cite{PHOTOS}.
Simulation also takes into account variations of the detector configuration 
and data-taking conditions.

\section{Event selection} \label{chapt2}

The event selection is performed in two steps. 
First we require the presence of at least two oppositely charged muon and two oppositely charged pion candidates.
Charged tracks must originate from a cylindrical region of length $\pm2.5$~cm along the $z$ axis (opposite the positron beam) and radius $2$~cm in the transverse plane, centered on the $e^+e^-$ interaction point. 
Muon candidates are identified with a requirement on a likelihood ratio $\mathcal{P}_{\mu}=\frac{\mathcal{L}_{\mu}}{\mathcal{L}_{\mu}+\mathcal{L}_{\pi}+\mathcal{L}_{K}}>0.1$ (efficiency is $\approx 99.9\%$), where the likelihood $\mathcal{L}_{i}$, with $i= \mu,\pi,K$, is assigned based on the range of the particle extrapolated from the CDC through KLM and on deviation of hits from extrapolated track \cite{Abashian:2002bd}.
Every charged particle that is not muon or not well identified as an electron 
($\mathcal{P}_e<0.99$) is considered as a charged pion candidate, where $\mathcal{P}_e$ is a similar likelihood ratio based on CDC, ACC, and ECL information \cite{Hanagaki:2001fz}. 
Additionally, we require dimuon invariant mass $M_{\mu\mu}=\sqrt{(P_{\mu^+}+P_{\mu^-})^2}$ to be in the range from 8 GeV/$c^2$ to 12 GeV/$c^2$ and dipion invariant mass $M_{\pi\pi}=\sqrt{(P_{\pi^+}+P_{\pi^-})^2}$ less than 4 GeV/$c^2$, where $P_i$ is the reconstructed four-momentum of a particle $i$. 
At this stage no requirements on photon candidates are applied.

Final-state-specific requirements are applied at the second stage. 
The following set of selection variables are common to all processes: 
the angle $\Psi$ between the total momentum of the photons and the total momentum 
of the charged particles in the CM frame, the invariant mass of the muon pair $M_{\mu\mu}$ (corresponding to the $\Upsilon(\rm 1S,2S)$), and the total reconstructed energy of the final-state particles, $E_{\rm tot}$. 
These variables are used to select exclusive decay chains that result in the same final states $\mu^+\mu^-\pi^+\pi^-\gamma(\gamma)$.

The signal region for $\Upsilon(\rm 1S)\to\mu^+\mu^-$ is defined to be $9.235\text{ GeV}/c^2 < M_{\mu\mu} <9.685\text{ GeV}/c^2$ and that for $\Upsilon(\rm 2S)\to\mu^+\mu^-$ is $9.76\text{ GeV}/c^2 < M_{\mu\mu} <10.28\text{ GeV}/c^2$.
Four-momentum conservation requires the angle $\Psi$ to be equal to 
$\pi$ radian; however, it can deviate even for true candidates due to 
finite momentum and energy resolutions.
For the $\Upsilon(\rm 2S)\eta[3\pi]$ mode this effect results in a less strict requirement on the angle $\Psi$ due to the low momentum of the $\pi^0$.
Selection criteria for the angle $\Psi$ are listed in Table \ref{tab:selection_criteria} for all modes.
In case of multiple decay candidates, usually due to additional photons in the event from background processes, the $\mu\mu\pi\pi\gamma(\gamma)$ combination with $\Psi$ closest to $\pi$ radian is chosen as the best candidate.
According to the simulation, fraction of multiple candidate events is about $24\%$ for the $\Upsilon(\rm 2S)\eta[3\pi]$ mode and ranges from $3\%$ to $12\%$ for other modes.
Finally, $E_{\rm tot}$ is calculated as
\begin{equation}
    E_{\rm tot}=E_{\pi\pi\gamma(\gamma)}+\sqrt{M_{\Upsilon(\rm 1S,2S)}^2+\vec{P}_{\mu\mu}^2},
\end{equation}
where instead of the reconstructed value of the $\mu^+\mu^-$-pair invariant mass the world-average mass of the $\Upsilon$ meson is used~\cite{PDG}.
This approach allows one to improve the $E_{\rm tot}$ resolution by
removing a contribution of the $M_{\mu\mu}$ resolution, whose value is about 
$50$~MeV/$c^2$ and comparable to the total contribution of all other 
terms in $E_{\rm tot}$.
Selection requirements on these common variables for all considered decay 
chains are summarized in Table~\ref{tab:selection_criteria}.
Additional criteria for selection of specific modes are described below. 

To reconstruct a neutral pion from the $\pi^0\to\gamma\gamma$ decay in the 
$\Upsilon(\rm 1S,2S)\eta[3\pi]$ modes, the invariant mass $M_{\gamma\gamma}$ 
should be in the signal range 
$110\text{ MeV}/c^2 < M_{\gamma\gamma} <155\text{ MeV}/c^2$, with resolution of $5.5$~MeV$/c^2$. 
For the $\Upsilon(\rm 1S)\eta^{\prime}[\pi\pi\eta]$ mode the $\eta$ meson is 
reconstructed from the $\eta\to\gamma\gamma$ decay with a signal range $450\text{ MeV}/c^2 < M_{\gamma\gamma} <625\text{ MeV}/c^2$, with resolution of $12.3$~MeV$/c^2$.
For the $\Upsilon(\rm 1S)\eta^{\prime}[\rho\gamma]$ mode the $\rho^0$ resonance is 
reconstructed from the $\rho^0\to\pi^+\pi^-$ decay with a signal range $450\text{ MeV}/c^2 < M_{\pi\pi} <950\text{ MeV}/c^2$, with root-mean-square of $56$~MeV$/c^2$.

For the two-body $\Upsilon(\rm 5S)\to\Upsilon(\rm 2S)\eta[\gamma\gamma]$ decay the 
produced $\eta$ meson is monochromatic, with a CM momentum equal to 
$615$~MeV/$c$.
Thus, photons produced in the $\eta\to\gamma\gamma$ decay have an energy in the CM frame distributed in the range from $105$~MeV to $715$~MeV.
We set a requirement on the minimum photon energy of $100$~MeV that significantly reduces combinatorial background and has virtually no effect on signal events.

For the $\Upsilon(\rm 2S)\eta[\gamma\gamma]$ final state the $\Upsilon(\rm 2S)$ meson 
is reconstructed via its decay chain $\Upsilon(\rm 2S)\to\Upsilon(\rm 1S)\pi^+\pi^-$ 
with $\Upsilon(\rm 1S)\to\mu^+\mu^-$.
The resolution of $M_{\mu\mu\pi\pi}$ is approximately $50$~MeV/$c^2$ and is dominated by 
the muon momentum resolution. 
To reduce this contribution we calculate the mass difference 
$M_{\mu\mu\pi\pi}-M_{\mu\mu}$, where the correlated contributions to resolution from the muon momentum measurement substantially cancel.
Due to the narrow width of the $\Upsilon(\rm 1S,2S)$ states, the $\delta M = M_{\mu\mu\pi\pi}-M_{\mu\mu}$ peak position corresponds to $\Delta M=M_{\Upsilon(\rm 2S)}-M_{\Upsilon(\rm 1S)} = 562$~MeV/$c^2$ while the resolution is approximately $4.6$~MeV/$c^2$, and a requirement of $|\delta M-\Delta M|<18$~MeV/$c^2$ is applied.

\begin{table*}[htb]
\footnotesize
\caption{Selection criteria and reconstruction efficiencies, where $M^{\rm rec}_{\pi\pi}=\sqrt{s + M^2_{\pi\pi} -2\sqrt{s}E_{\pi\pi}}$, $\delta M=M_{\mu\mu\pi\pi}-M_{\mu\mu}$, and $\Delta M_2=M_{\Upsilon(\rm 2S)}-M_{\Upsilon(\rm 1S)} = 562$~MeV/$c^2$, $\Delta M_3=M_{\Upsilon(\rm 3S)}-M_{\Upsilon(\rm 1S)} = 894$~MeV/$c^2$.}
    \label{tab:selection_criteria}
\begin{tabular}
 {@{\hspace{0.4cm}}l@{\hspace{0.2cm}}  @{\hspace{0.4cm}}c@{\hspace{0.4cm}} @{\hspace{0.4cm}}c@{\hspace{0.4cm}} @{\hspace{0.4cm}}c@{\hspace{0.4cm}} @{\hspace{0.4cm}}c@{\hspace{0.2cm}} @{\hspace{0.2cm}}c@{\hspace{0.1cm}}}
\hline \hline

Criterion & $\Upsilon(\rm 2S)\eta[3\pi]$ & $\Upsilon(\rm 2S)\eta[\gamma\gamma]$  &   $\Upsilon(\rm 1S)\eta[3\pi]$ & $\Upsilon(\rm 1S)\eta^{\prime}[\pi\pi\eta]$ & $\Upsilon(\rm 1S)\eta^{\prime}[\rho\gamma]$\\
        \hline
        $M_{\mu\mu}$, GeV/$c^2$  & $[9.76,~10.28]$ & $[9.235,~9.685]$ & $[9.235,~9.685]$ & $[9.235,~9.685]$ & $[9.235,~9.685]$  \\
        $\Psi$, rad. &   $\geq 2$  & $\geq 2.8$& $\geq 2.7$ & $\geq 2.8$ & $\geq 2.5$ \\
        $E_{\rm tot}$, GeV &  $[10.775,~10.92]$ & $[10.80,~10.955]$ & $[10.75,~10.94]$ & $[10.75,~10.94]$ & $[10.75,~10.94]$\\
       \hline
       $M_{\gamma\gamma}$, MeV/$c^2$ &  $[110,~155]$ & -- & $[110,~155]$ & $[450,~625]$ & --\\
       $\delta M$, MeV/$c^2$ &  -- & $|\delta M-\Delta M_2|<18$  & $|\delta M-\Delta M_2|>10$ & $|\delta M-\Delta M_2|>10$ & $|\delta M-\Delta M_{2,3}|>10$ \\
       $\alpha_{\pi\pi}$, rad.  &$\geq 0.3$ &  -- &  $\geq 0.18$ & -- & --  \\
       $E^*_\gamma$, MeV &  --  & $>100$   & -- & -- & $>80$ \\
       $M_{\pi\pi}$, MeV/$c^2$ & -- & -- & -- & -- & $[450,~950]$\\
       $M^{\rm rec}_{\pi\pi}$, MeV/$c^2$ & -- & -- & -- & -- & $|M^{\rm rec}_{\pi\pi}-M_{\Upsilon(\rm 2S)}|>20$\\
       \hline
         $\varepsilon$, $\%$ & $10.25 \pm 0.03$ & $20.73\pm 0.04$ &  $17.02\pm 0.03$ & $13.35\pm 0.03$ & $29.25\pm 0.05$ \\
\hline \hline
\end{tabular}
\end{table*}

The signal distribution for all modes is the $M_{\eta^{(\prime)}}$ ($M_{\gamma\gamma}$, $M_{\pi\pi\gamma\gamma}$ or $M_{\pi\pi\gamma}$) invariant mass, it having no peaking background (see Section \ref{sect_bkg}).
The MC signal distribution is fitted by a sum of the Crystal Ball function \cite{Skwarnicki:1986xj} and a Gaussian.
The reconstruction efficiency $\varepsilon$ is then determined as $N_{\rm det}/N_{\rm gen}$, where $N_{\rm det}$ is the integral of the fitted function and $N_{\rm gen} = 10^6$. 
Results are summarized in Table~\ref{tab:selection_criteria}.


\section{Study of the expected background} \label{sect_bkg} 

The most relevant background to this analysis comes from transitions between other bottomonium states with emission of an $\eta^{(\prime)}$ meson. These decays have an $\eta^{(\prime)}$ invariant mass distribution identical to our signal modes.

Due to the $\eta^{\prime}$ mass and parity considerations, the $\eta^{\prime}$ meson can originate only from the $\Upsilon(\rm 5S)\to\Upsilon(\rm 1S)\eta^{\prime}$ decay or from $\Upsilon(\rm 5S)\to\chi_{bJ}(1P)\eta^{\prime}\gamma$ decays with a subsequent radiative decay of $\rm \chi_{bJ}(\rm 1P)\to\Upsilon(\rm 1S)\gamma$. 
The former is our signal and the latter is suppressed by the presence of an additional photon.

In contrast, the $\eta$ meson can also originate from $\Upsilon(\rm 5S)\to\Upsilon(\rm 1D)\eta$ \cite{Tamponi:2018cuf} and $\Upsilon(\rm 5S)\to\Upsilon(\rm 2S,3S)X$ followed by $\Upsilon(\rm 2S,3S)\to\Upsilon(\rm 1S)\eta$ decays.
For the $\Upsilon(\rm 5S)\to\Upsilon(\rm 1D)\eta$ decay, the most relevant channels are those with $\Upsilon(\rm 1D) \to \chi_{bJ}\gamma \to \Upsilon(\rm 1S)\gamma\gamma$ and $\Upsilon(\rm 1D)\to\Upsilon(\rm 1S)\pi^+\pi^-$ decays.
However, the first decay has two extra photons and is suppressed by the requirement on $E_{\rm tot}$.
The second decay might produce a correct set of final-state particles (with $\eta\to\gamma\gamma$), but is significantly suppressed by the requirement on $M_{\mu\mu\pi\pi}-M_{\mu\mu}$: for the $\Upsilon(\rm 1D)\to\Upsilon(\rm 1S)\pi^+\pi^-$ decay, this variable peaks at approximately $140$~MeV/$c^2$ higher than for the $\Upsilon(\rm 2S)\to\Upsilon(\rm 1S)\pi^+\pi^-$ with a resolution of about $5$~MeV/$c^2$. Therefore, the $\Upsilon(\rm 1D)\to\Upsilon(\rm 1S)\pi^+\pi^-$ signal is completely eliminated.

$\Upsilon(\rm 2S,3S)$ mesons mostly originate from $\Upsilon(\rm 5S)\to\Upsilon(\rm 2S,3S)\pi^+\pi^-$ decays. 
With subsequent $\Upsilon(\rm 2S,3S)\to\Upsilon(\rm 1S)\eta$ and $\eta\to\gamma\gamma$ decays these channels produce the same set of final-state particles and the same signal distribution as the $\Upsilon(\rm 2S)\eta[\gamma\gamma]$ mode. 
However, branching fractions $\mathcal{B}(\Upsilon(\rm 2S,3S)\to\Upsilon(\rm 1S)\eta)$ are small 
and with the current integrated luminosity the expected number of $\eta$ mesons produced by this mechanism is estimated to be $2$ for the $\Upsilon(\rm 2S)$ and less than $1$ for the $\Upsilon(\rm 3S)$. 
Contributions from these decays are also strongly suppressed to negligible level by the requirement on $M_{\mu\mu\pi\pi}-M_{\mu\mu}$; its mean value deviates from the $\Upsilon(\rm 2S)\to\Upsilon(\rm 1S)\pi^+\pi^-$ signal window by $280$~MeV/$c^2$ and $50$~MeV/$c^2$ for $\Upsilon(\rm 5S)\to\Upsilon(\rm 2S)\pi^+\pi^-$ and $\Upsilon(\rm 5S)\to\Upsilon(\rm 3S)\pi^+\pi^-$ correspondingly.

A possible source of background for $\Upsilon(\rm 5S)\to\Upsilon(\rm 2S)\eta$ is the decay itself with $\Upsilon(\rm 2S)\to\Upsilon(\rm 1S)\eta$, where $\eta\to\pi^+\pi^-\pi^0$ or $\eta\to\pi^+\pi^-\gamma$. This final state is similar to the $\Upsilon(\rm 2S)\eta[\gamma\gamma]$ mode when a soft photon or $\pi^0$ is undetected. 
Nevertheless, this background is suppressed to negligible level by the intermediate branching-fraction ratio $\frac{\mathcal{B}(\Upsilon(\rm 2S)\to\Upsilon(\rm 1S)\eta)\times \mathcal{B}(\eta\to\pi^+\pi^-\pi^0(\gamma))}{\mathcal{B}(\Upsilon(\rm 2S)\to\Upsilon(\rm 1S)\pi^+\pi^-)}\sim 4\times 10^{-4}$ and requirements on $E_{\rm tot}$.

Crossfeed between the signal modes is a background that passes through the common selection criteria but does not produce peaks in the signal distributions.
For the $\Upsilon(\rm 1S)\eta[3\pi]$ and $\Upsilon(\rm 1S)\eta^{\prime}$ modes, there is such a background from the $\Upsilon(\rm 2S)\eta[\gamma\gamma]$ mode when $\Upsilon(\rm 2S)\to\Upsilon(\rm 1S)\pi^+\pi^-$.
To reduce this background for the $\Upsilon(\rm 1S)\eta^{(\prime)}$ mode, we require $|M_{\mu\mu\pi\pi}-M_{\mu\mu}-(M_{\Upsilon(\rm 2S)}-M_{\Upsilon(\rm 1S)})|>10$~MeV/$c^2$, which only slightly decreases the signal reconstruction efficiency and suppress this background to negligible level. 
The crossfeed between the $\Upsilon(\rm 1S)\eta[3\pi]$ and $\Upsilon(\rm 2S)\eta[3\pi]$ modes is efficiently removed by the common selection requirements.

Another significant part of the background is the non-peaking combinatorial 
background. 
To evaluate the expected level of this background, we used a set of MC events six times larger than that in data including the following processes: 
$e^+e^-\to$ $c\Bar{c}$, $u\Bar{u}$, $d\Bar{d}$, $s\Bar{s}$; $e^+e^-\to$ $B^{(*)}_s\Bar{B}^{(*)}_s$, $B^{(*)}\Bar{B}^{(*)}$; and known decays of the $\Upsilon(\rm 5S)$.
In addition, we performed simulation of $e^+e^-\to\tau^+\tau^-$ events with 
statistics equivalent to the integrated luminosity of our dataset.
The only events remaining after application of the selection criteria originate from $\Upsilon(\rm 5S)$ decays to final states containing bottomonium.
As an example, the dominant background to the $\Upsilon(\rm 2S)\eta[\gamma\gamma]$ comes from the $\Upsilon(\rm 2S)\pi^0\pi^0$ final state, which produces a broad peaking $M_{\gamma\gamma}$ distribution from $50$~MeV/$c^2$ to $850$~MeV/$c^2$ with a maximum near the signal $\eta$ peak position. 
To suppress this background we increased the requirement on the total reconstructed energy from $10.75$~GeV to $10.80$~GeV for the $\Upsilon(\rm 2S)\eta[\gamma\gamma]$ mode. 
This reduces the expected number of background events for the $\Upsilon(\rm 2S)\eta[\gamma\gamma]$ from $20$ to $5$ events and slightly decreases the detection efficiency.

For the $\Upsilon(\rm 1S)\eta^{\prime}[\rho\gamma]$ mode, the MC study predicts high background from the $\Upsilon(\rm 5S)\to\Upsilon(\rm 2S)\pi^+\pi^-$ decay, where $\Upsilon(\rm 2S)\to \mu^+\mu^-\gamma(\gamma)$.
To reduce this background we set a veto on recoil mass $M^{\rm rec}_{\pi\pi}$: $|M^{\rm rec}_{\pi\pi}-M_{\Upsilon(\rm 2S)}|>20$~MeV/$c^2$.
The MC study also predicts background contributions from decays with the $\Upsilon(\rm 2S,3S)\to\Upsilon(\rm 1S)[\mu^+\mu^-]\pi^+\pi^-$ intermediate transition.
Therefore, we reduced this background in the same way as for the $\Upsilon(\rm 1S)\eta[3\pi]$ and $\Upsilon(\rm 1S)\eta^{\prime}[\pi\pi\eta]$ modes by setting vetoes $|M_{\mu\mu\pi\pi}-M_{\mu\mu}-(M_{\Upsilon(\rm 2S,3S)}-M_{\Upsilon(\rm 1S)})|>10$~MeV/$c^2$.
Moreover, there are no requirements on photons except the general one from the four-momentum conservation, therefore we expect low-energy background photons.
To suppress this background we set a minimum photon energy in the CM frame of $E^*_{\gamma}>80$~MeV.
It does reduce reconstruction efficiency by factor of $1.12$, but also greatly reduces background.

We find that all these sources predicted by the background MC account for 
less than $30\%$ of the observed background in the side-band data.
The rest of the background comes presumably from QED processes that in general 
have much higher cross sections, $\emph{e.g.}$
processes like $e^+e^-\to\mu^+\mu^-e^+e^-$.
This $e^+e^-$ pair could be reconstructed as a pair of collinear pions.
A selection requirement on the opening angle between two charged pion 
candidates of $\alpha_{\pi\pi} >0.18$ radian for the $\Upsilon(\rm 1S)\eta[3\pi]$ mode and of $\alpha_{\pi\pi}>0.3$ radian for the $\Upsilon(\rm 2S)\eta[3\pi]$ mode reduces this background substantially.

Finally, we tested for possible background from nonresonant $e^+e^-\to\mu^+\mu^-\eta^{(\prime)}$ decays using experimental data with a requirement on $M_{\mu\mu}$ shifted to a range from $8$~GeV/$c^2$ to $9$~GeV/$c^2$ that is lower than the ground bottomonium state.  
No evidence for such processes was observed.

\section{Data analysis} \label{chapt3}

\subsection{Cross section at the $\Upsilon$(5S) resonance} \label{sect_on_res}
\begin{figure}[!t]
  \centering
  \includegraphics[width=8.6cm]{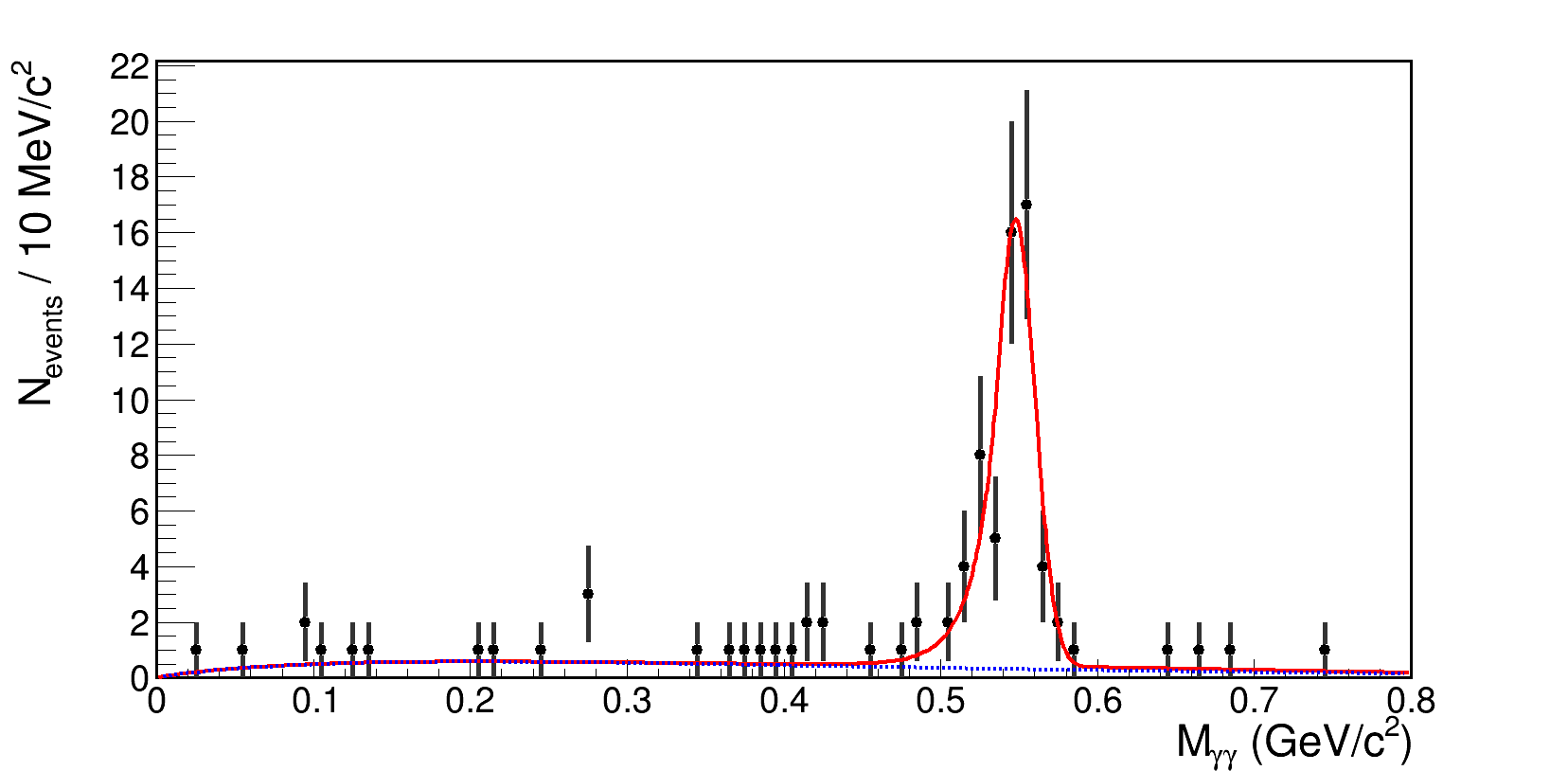} \hfill
  \includegraphics[width=8.6cm]{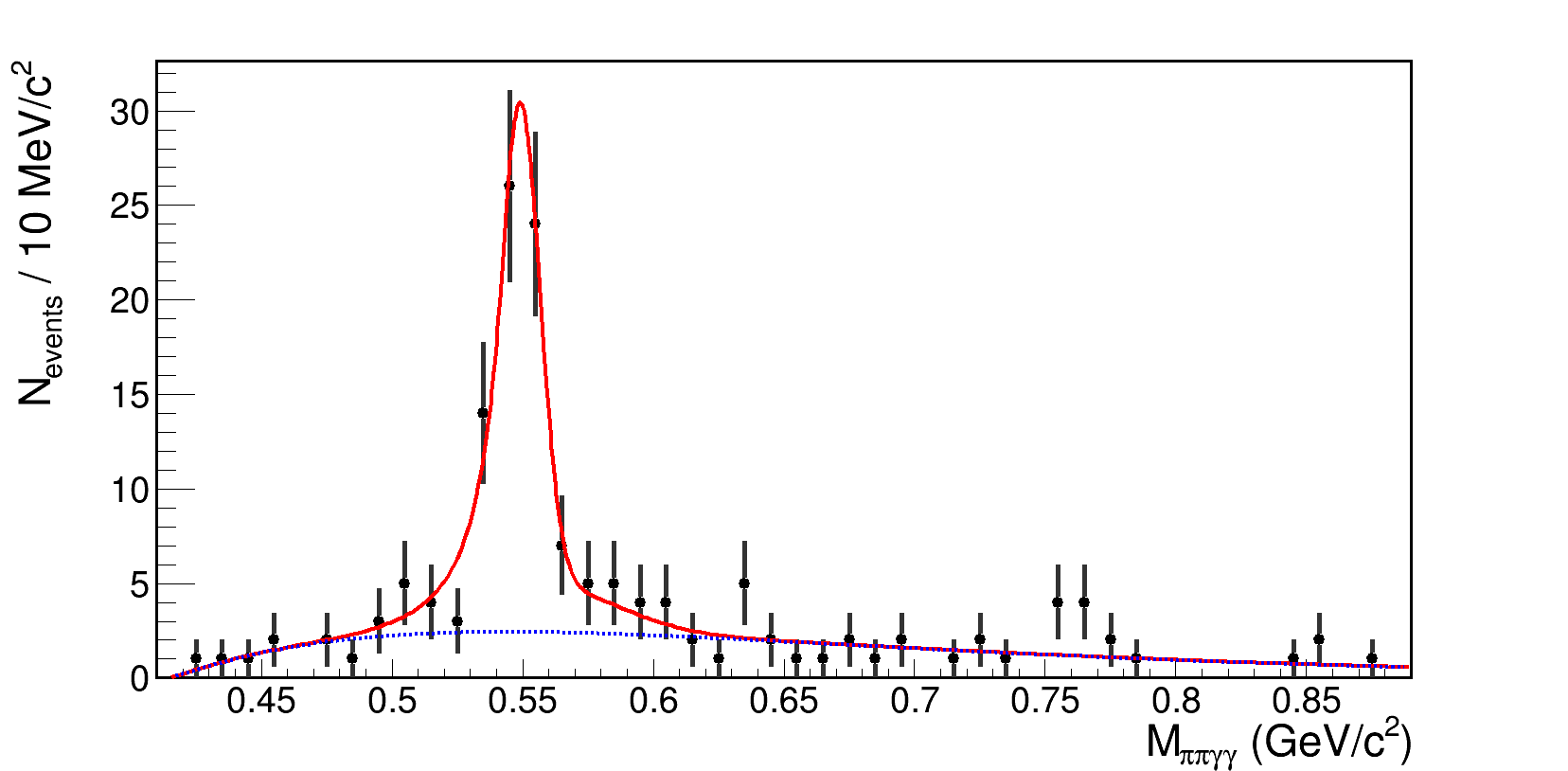} \\
  \includegraphics[width=8.6cm]{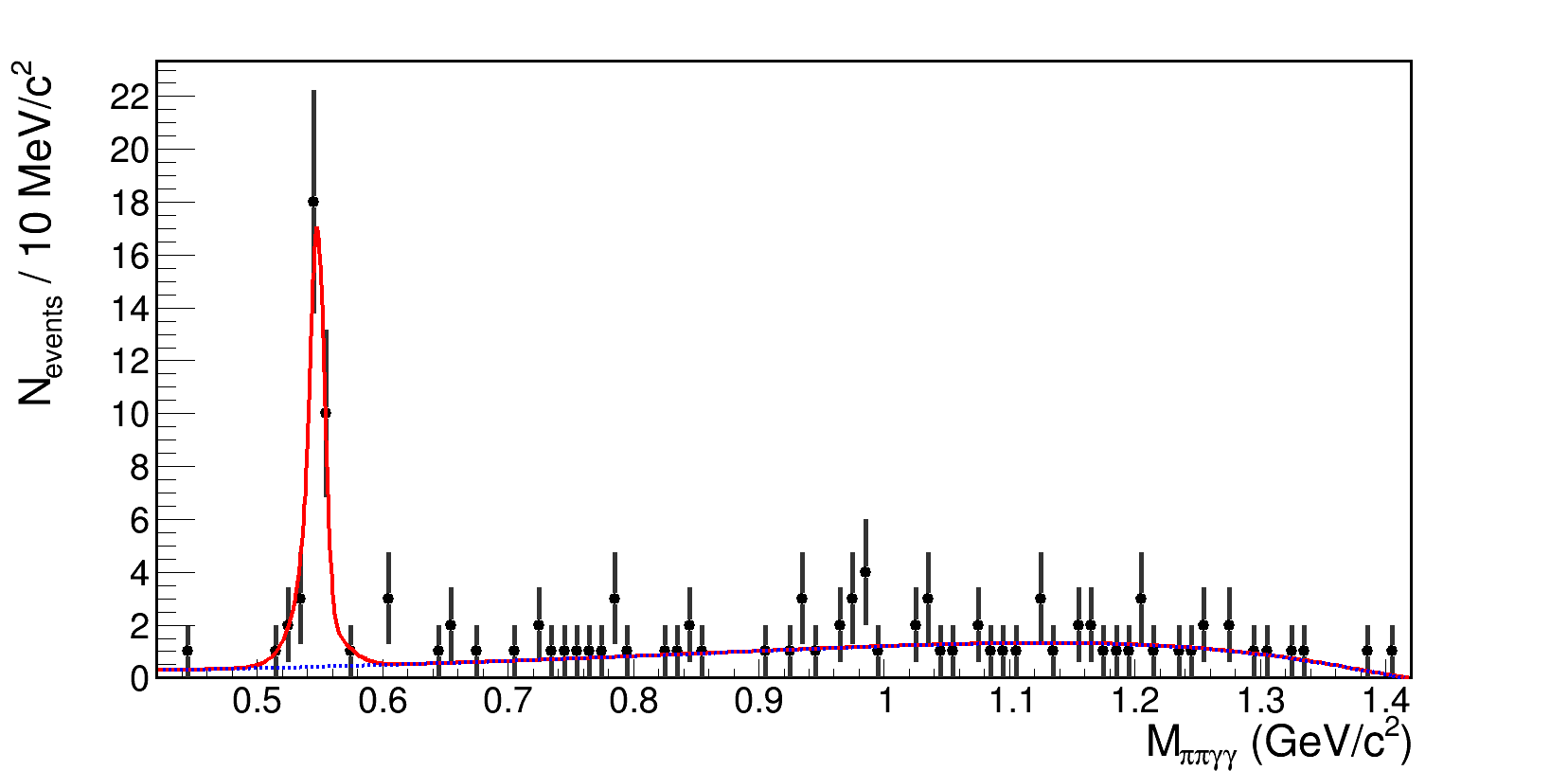} 
 
   \caption{The experimental signal $M_{\eta}$ ($M_{\gamma\gamma}$ or $M_{\pi\pi\gamma\gamma}$) distribution for $\Upsilon(\rm 2S)\eta[\gamma\gamma]$ (a), $\Upsilon(\rm 2S)\eta[3\pi]$ (b) and  $\Upsilon(\rm 1S)\eta[3\pi]$ (c) fitted to the sum of the MC signal function and background function $(x-p_1)^{p_2}e^{p_3x}$. Data are shown as points, the solid red line shows the best fit to the data, and the dashed blue line shows the background contribution.}   
    \label{fig:meta_exp_fit}
\end{figure}

\begin{figure}[!t]
  \centering
  \includegraphics[width=8.6cm]{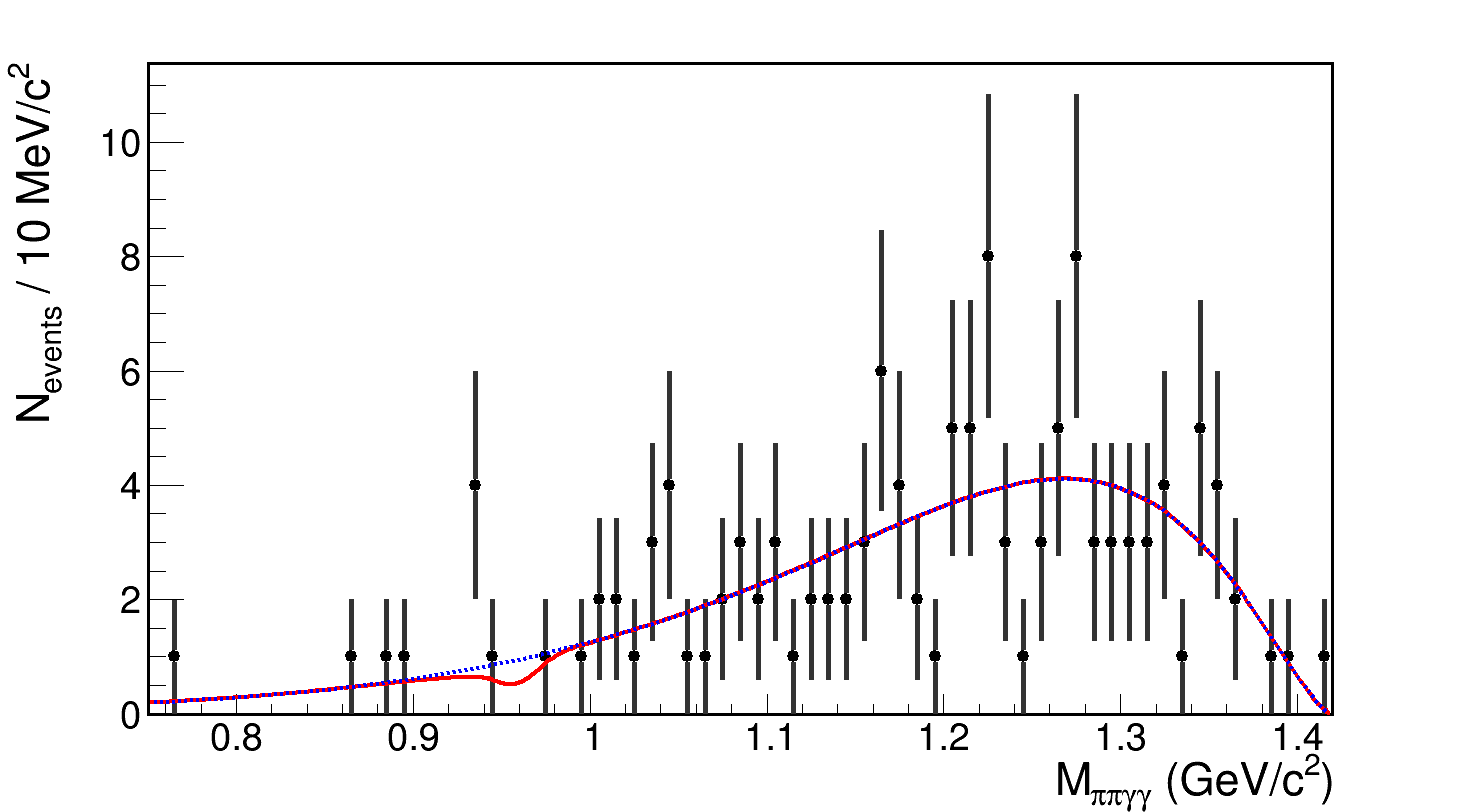} \hfill 
  \includegraphics[width=8.6cm]{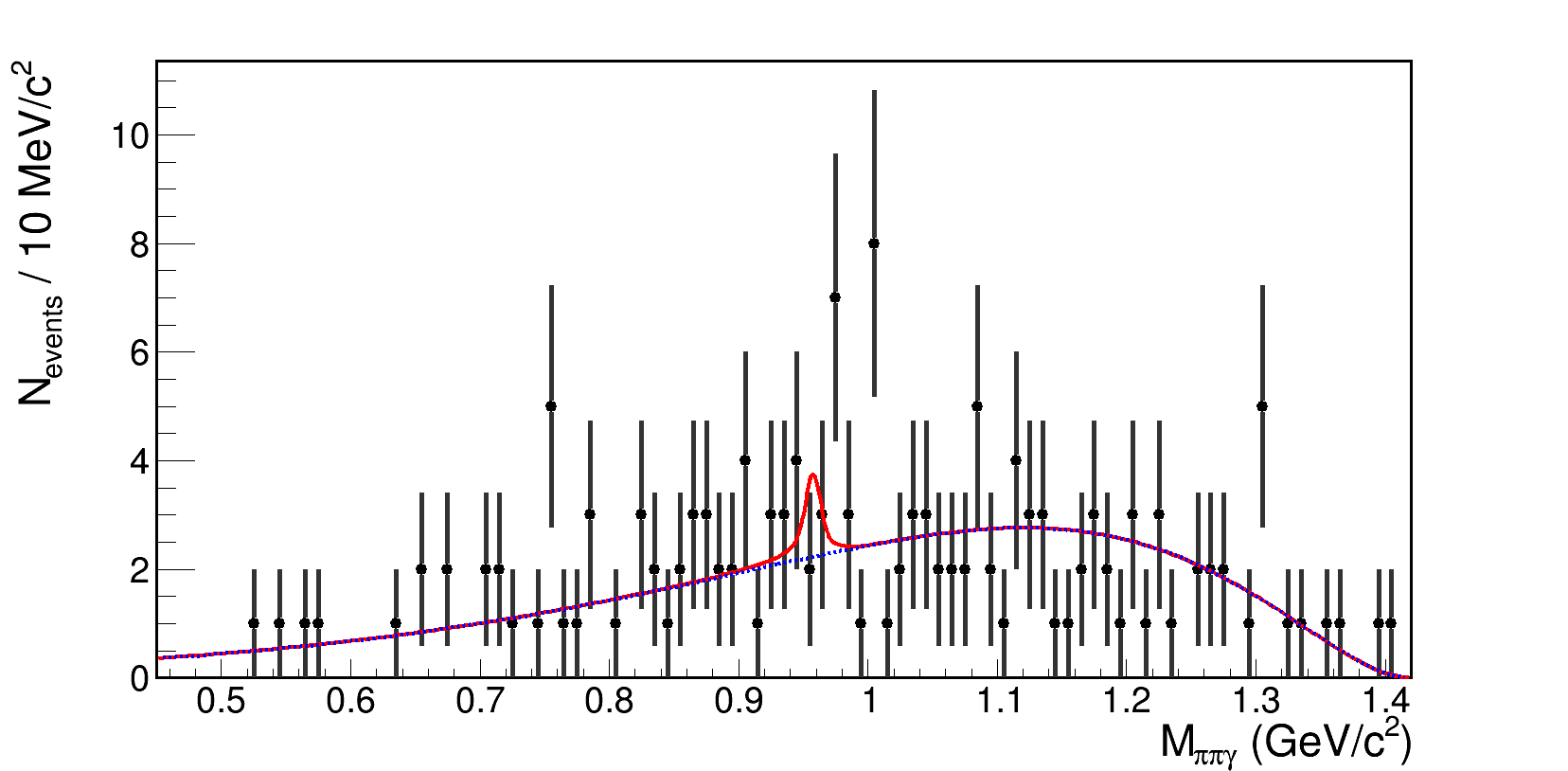}
   \caption{The experimental signal $M_{\eta^{\prime}}$ ($M_{\pi\pi\gamma\gamma}$ or $M_{\pi\pi\gamma}$) distribution for $\Upsilon(\rm 1S)\eta^{\prime}[\pi\pi\eta]$ (a) and $\Upsilon(\rm 1S)\eta^{\prime}[\rho\gamma]$ (b) fitted to the sum of the MC signal function and background function $(x-p_1)^{p_2}e^{p_3x}$. Data are shown as points, the solid red line shows the best fit to the data, and the dashed blue line shows the background contribution.}   
    \label{fig:meta2_exp_fit}
\end{figure}

The signal yield is determined from a binned maximum likelihood fit to the invariant mass 
$M_{\eta^{(\prime)}}$ ($M_{\gamma\gamma}$, $M_{\pi\pi\gamma\gamma}$ or $M_{\pi\pi\gamma}$) distribution 
(Fig.~\ref{fig:meta_exp_fit},~\ref{fig:meta2_exp_fit}), with the fitting function being the sum 
of the signal function and a background function $(x-p_1)^{p_2}e^{p_3x}$, where $p_1$, $p_2$, $p_3$ are floating parameters.
All parameters of the signal function, except its normalization factor and the Crystal Ball peak position, are fixed to the values determined from the fit to the MC distribution, with a relative distance between Crystal Ball and Gaussian peaks being fixed.

The visible cross section is
\begin{equation}
	\label{vis_section}
    \sigma_{\rm vis} = \frac{N_{\rm sig}}{L \mathcal{B}\varepsilon},
\end{equation}
where $N_{\rm sig}$ is the fitted signal yield, $L$ is the integrated luminosity, 
$\mathcal{B}$ is the product of the intermediate branching fractions for the process, 
and $\varepsilon$ is the reconstruction efficiency.

\begin{figure*}[!t]
  \centering
  \includegraphics[width=0.32\textwidth]{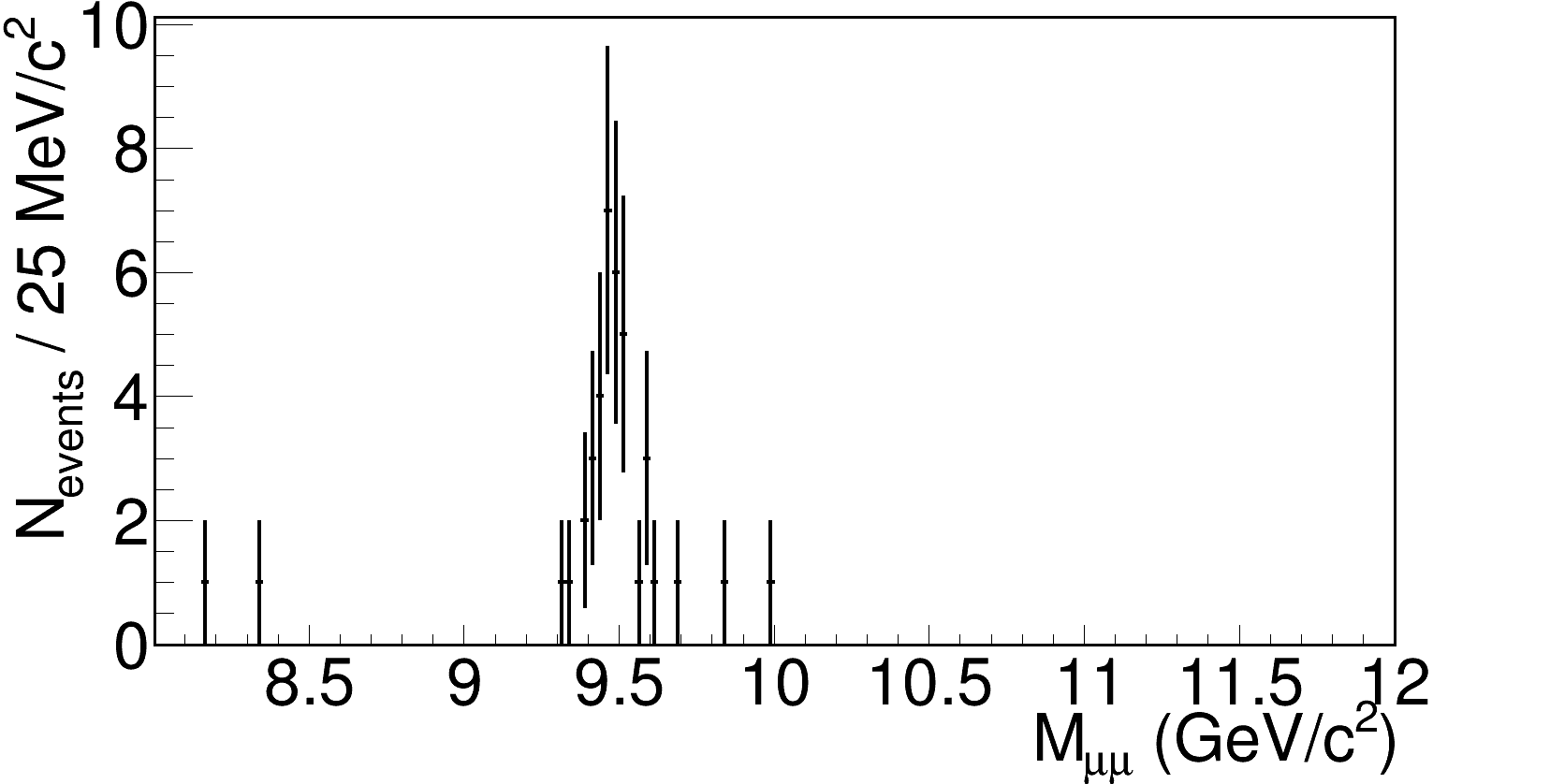} \hfill
  \includegraphics[width=0.32\textwidth]{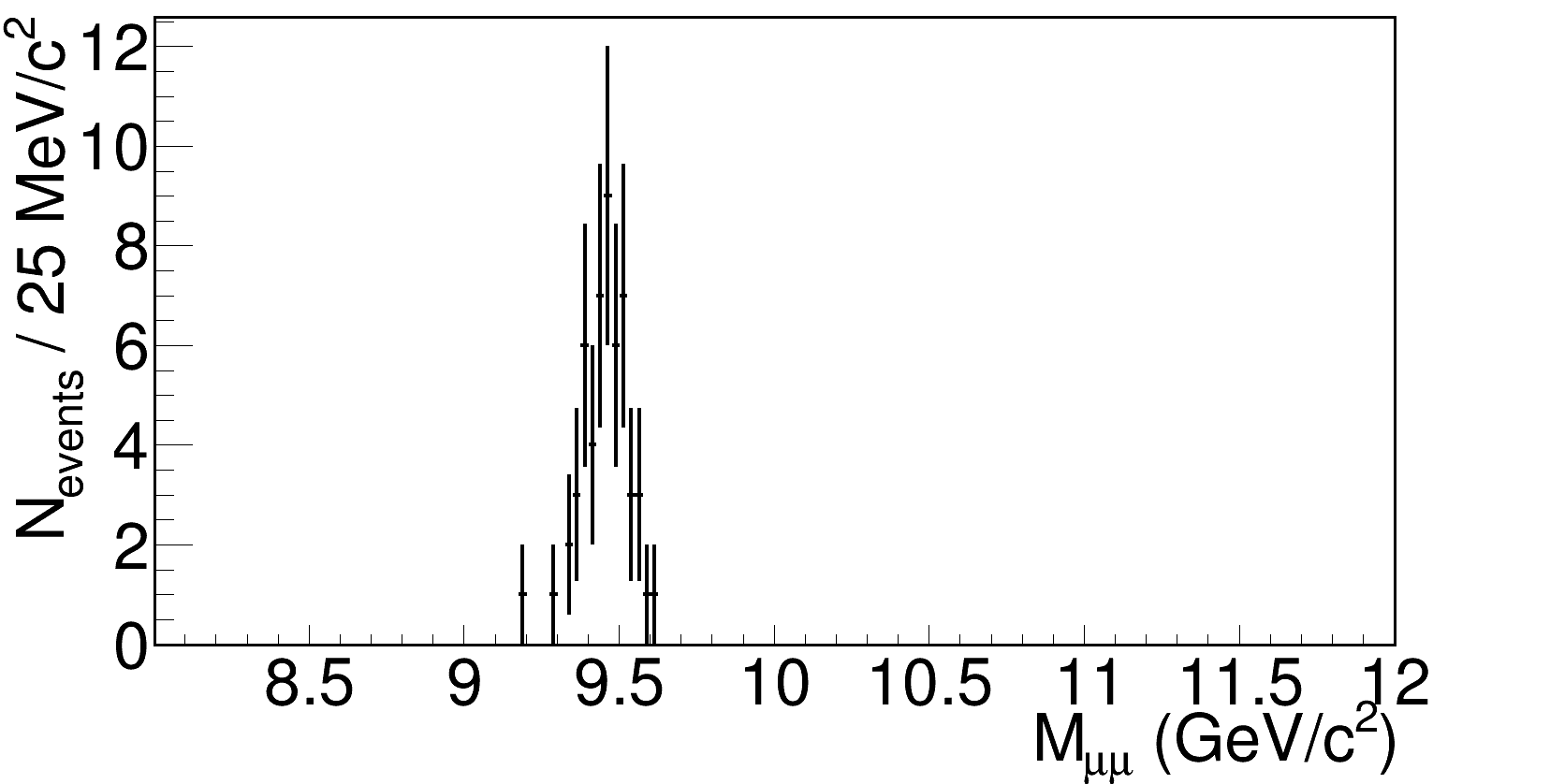} \hfill
  \includegraphics[width=0.32\textwidth]{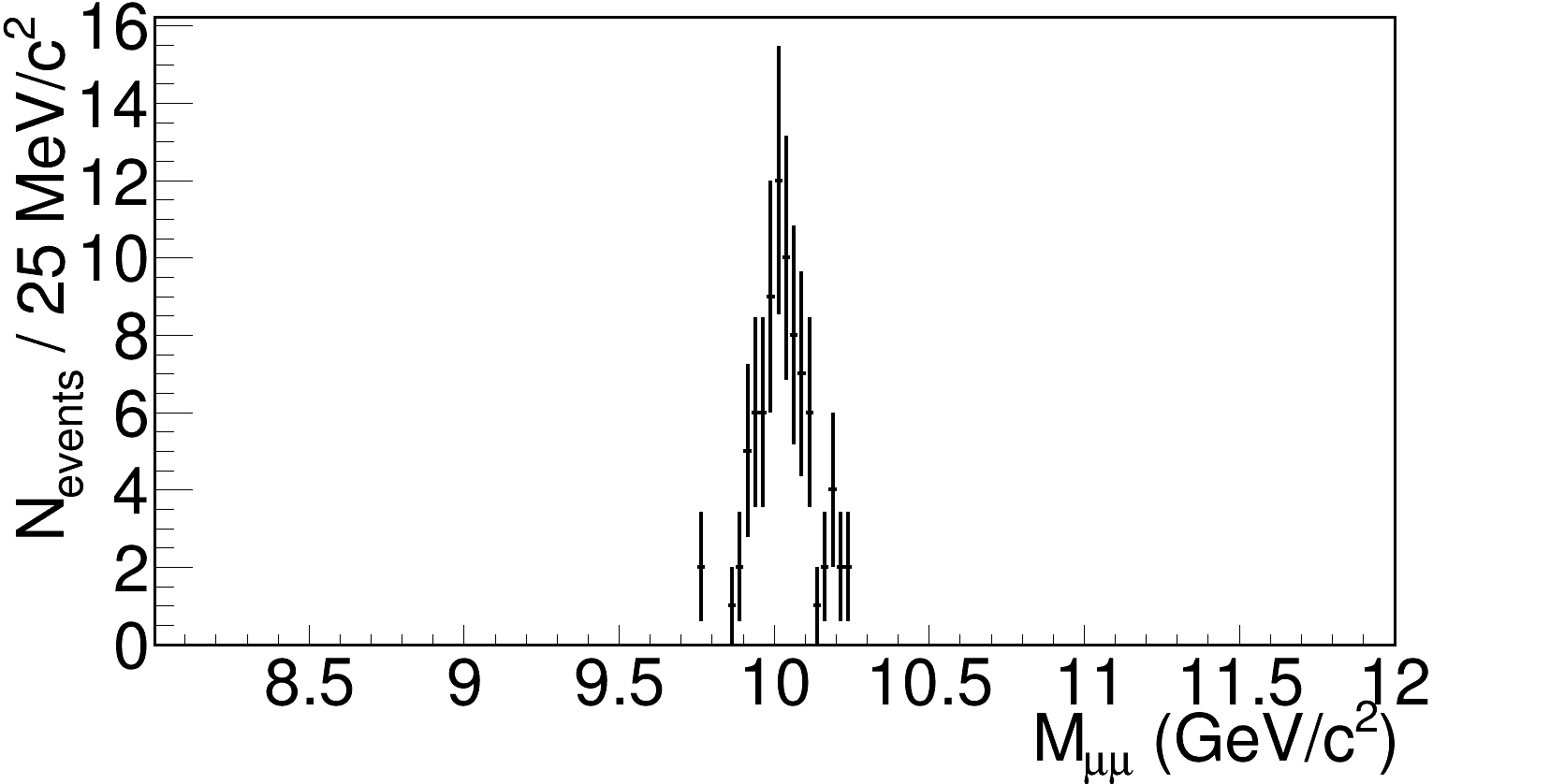} 
 
   \caption{The experimental $M_{\mu\mu}$ distribution with requirement $520<M_{\eta}<580$~MeV for $\Upsilon(\rm 1S)\eta[3\pi]$~(a), $\Upsilon(\rm 2S)\eta[\gamma\gamma]$~(b), and $\Upsilon(\rm 2S)\eta[3\pi]$~(c) modes. No requirement on $M_{\mu\mu}$ is applied.}   
    \label{fig:mumu_exp}
\end{figure*}

For the $\Upsilon(\rm 2S)\eta[\gamma\gamma]$, $\Upsilon(\rm 2S)\eta[3\pi]$, and $\Upsilon(\rm 1S)\eta[3\pi]$ modes we evaluate the signal significance in standard deviations as $\sqrt{2 \log{[\mathcal{L}(N)/\mathcal{L}(0)}]}$,
where $\mathcal{L}(N)/\mathcal{L}(0)$ is the ratio between the likelihood values for a fit that includes a signal yield $N$ and a fit with a background hypothesis only.
The calculated signals significance are $12.8\sigma$, $10.5\sigma$, and $10.2\sigma$ respectively.
Thus, we report the first observation of the $e^+e^- \to \Upsilon(\rm 2S)\eta$ process in both modes with the quadratically combined significance of $16.5\sigma$, and the first observation of the $e^+e^- \to \Upsilon(\rm 1S)\eta$ process with the significance of $10.2\sigma$ at $\sqrt{s}=10.866$~GeV.
Setting requirement $520<M_{\eta}<580$~MeV, we also confirm that there are clear peaks in $M_{\mu\mu}$ distributions (Fig. ~\ref{fig:mumu_exp}) consistent with corresponding $\Upsilon(\rm 1S,2S)\to\mu^+\mu^-$ for these modes.

For the $\Upsilon(\rm 1S)\eta^{\prime}[\pi\pi\eta]$ and the $\Upsilon(\rm 1S)\eta^{\prime}[\rho\gamma]$ modes the signal yield is $N_{\rm sig} = -1.76 \pm 3.30$ and $N_{\rm sig} = 3.30 \pm 4.41$ correspondingly; therefore, only upper limits are set using a pseudo-experiment method.
Within the method we simulate $10^5$ trials, each having its 
own number of events sampled with the Poisson distribution having a mean
of total events in the experimental signal distribution.
For each event the value of $M_{\pi\pi\gamma(\gamma)}$ is sampled using the data
background distribution.
Then the obtained signal $M_{\pi\pi\gamma(\gamma)}$ distribution for each trial is fitted with the same procedure as the data and the obtained signal yield is recorded.
We determine a confidence level ($CL$) on the upper limit as a ratio of the number of trials, which gave the signal yield from $0$ to $N_{\rm sig}$, to the total number of trials with $N_{\rm sig}$ above $0$.
As a result, the $90\%~CL$ upper limits for the $\Upsilon(\rm 1S)\eta^{\prime}[\pi\pi\eta]$ and $\Upsilon(\rm 1S)\eta^{\prime}[\rho\gamma]$ modes are  $N_{\rm sig}=5.2$ and $N_{\rm sig}=5.6$, respectively.

Table~\ref{tab:exp_ndet} shows the signal yield, calculated visible cross 
section for all modes, and peak positions for the $\eta$ meson, which is 
consistent with the world-average value $M_{\eta}=547.86 \pm 0.02$~MeV/$c^2$ \cite{PDG}
within statistical uncertainty.

\subsection{Cross section outside of $\Upsilon(\rm 5S)$} \label{sect_off_res}

It is necessary to study the cross section behavior of the processes below 
the $\Upsilon(\rm 5S)$ to calculate radiative corrections.
For that purpose we use $21~\text{fb}^{-1}$ of the scan data collected in the energy range 
from $10.63$~GeV to $11.02$ GeV.
We group the scan data into three ranges: $10.63$~GeV -- $10.77$~GeV (below $\Upsilon(\rm 5S)$), $10.83$~GeV -- $10.91$~GeV (at $\Upsilon(\rm 5S)$) and $10.93$~GeV -- $11.02$~GeV (at $\Upsilon(\rm 6S)$).
These data sets are analysed in the same way as the main one except for the 
requirement on $E_{\rm tot}$, which is shifted to the corresponding energy.
Analysis shows (Table~\ref{tab:exp_ndet_off}) that there are no signal events 
below the $\Upsilon(\rm 5S)$ resonance except one event for the $\Upsilon(\rm 2S)\eta[3\pi]$ mode.
Thus, we set upper limits for each mode $N_{\rm sig}<1$ corresponding to a $CL$ of $63\%$.

For $\Upsilon(\rm 1S)\eta[3\pi]$, $\Upsilon(\rm 1S)\eta^{\prime}[\pi\pi\eta]$ and $\Upsilon(\rm 1S)\eta^{\prime}[\rho\gamma]$ modes, the upper 
limits are higher than values measured at the $\Upsilon(\rm 5S)$ resonance, while 
for the $e^+e^-\to \Upsilon(\rm 2S)\eta $ process the upper limit does not 
contradict resonance production of the final state.
Other examples of resonance production are similar processes 
$e^+e^-\to\Upsilon(\rm 1S,2S,3S)\pi^+\pi^-$~\cite{Ypipi_scan, Ypipi_scan_new} and $e^+e^-\to h_b(\rm 1P,2P)\pi^+\pi^-$~\cite{hb_scan}.
Since the scan data results do not contradict this assumption, we use a 
resonance model to calculate a radiative correction for all modes as is 
described, for example, in~\cite{Benayoun:1999hm}, neglecting the possible energy dependence of the resonance width.
For this calculation, the following $\Upsilon(\rm 5S)$ parameters are used: $M_{\Upsilon(\rm 5S)}=10885.2$~MeV/$c^2$, $\Gamma_{\Upsilon(\rm 5S)}=37$~MeV.
The calculated radiative correction $1+\delta$ varies from $0.624$ to $0.628$ for different modes.
This correction is used to calculate Born cross section ($\sigma_{\rm B}$) as
\begin{equation}
    \label{born crossection}
    \sigma_{\rm B} = \sigma_{\rm vis}\frac{|1-\Pi|^2}{1+\delta},
\end{equation}
where $|1-\Pi|^2=0.929$ is the vacuum-polarization factor \cite{Tamponi:2018cuf, Actis:2010gg}.

\begin{table*}[htb]
\caption{Signal yield, visible cross section and $M_{\eta^{(\prime)}}$ peak position for all modes at $\sqrt{s}=10.866$ GeV. The uncertainty is statistical only.}
    \label{tab:exp_ndet}
\begin{tabular}
 {@{\hspace{0.5cm}}l@{\hspace{0.5cm}}  @{\hspace{0.5cm}}c@{\hspace{0.5cm}}  @{\hspace{0.5cm}}c@{\hspace{0.5cm}}  @{\hspace{0.5cm}}c@{\hspace{0.5cm}}}
\hline \hline
        Mode & $N_{\rm sig}$ & $\sigma_{\rm vis}$, pb & $M_{\eta^{(\prime)}}$ exp., MeV/$c^2$  \\
        \hline
        $\Upsilon(\rm 2S)\eta[\gamma\gamma]$ & $59.5\pm 8.3$ & $1.39\pm 0.19$ &  $547.8 \pm 2.0$ \\
        $\Upsilon(\rm 2S)\eta[3\pi]$ & $73.8\pm 10.7$ & $1.39\pm 0.20$ &   $549.1 \pm 1.5$ \\ 
        $\Upsilon(\rm 1S)\eta[3\pi]$ & $32.6\pm 5.9$ & $0.29\pm 0.05$ &    $547.9 \pm 1.3$ \\
        $\Upsilon(\rm 1S)\eta^{\prime}[\pi\pi\eta]$ & $<5.2$, $CL=90\%$ & $<0.080$, $CL=90\%$ & -- \\
        $\Upsilon(\rm 1S)\eta^{\prime}[\rho\gamma]$ & $<5.6$, $CL=90\%$ & $<0.022$, $CL=90\%$ & -- \\ 
\hline \hline
\end{tabular}
\end{table*}

\begin{table*}[htb!]
\caption{Results of the scan data analysis and comparison of the averaged upper limits on cross section below $\Upsilon(\rm 5S)$ with results from the main data set. $N_{\rm sig}$ is the signal yield and 
$N_{\rm tot}$ is the total number of events in the signal distribution.}
\label{tab:exp_ndet_off}
\begin{tabular}
 {@{\hspace{0.4cm}}l@{\hspace{0.4cm}}  @{\hspace{0.2cm}}c@{\hspace{0.2cm}}  @{\hspace{0.4cm}}c@{\hspace{0.4cm}}  @{\hspace{0.4cm}}c@{\hspace{0.4cm}} @{\hspace{0.4cm}}c@{\hspace{0.4cm}}  @{\hspace{0.4cm}}|c@{\hspace{0.4cm}} @{\hspace{0.4cm}}c@{\hspace{0.4cm}} }
\hline \hline
     Mode & $\sqrt{s}$ range, GeV & $L, ~\text{fb}^{-1}$  & $N_{\rm sig}$ & $N_{\rm tot}$ & $\sigma_{\rm vis}$ below $\Upsilon(\rm 5S)$, pb & $\sigma_{\rm vis}$ at $\Upsilon(\rm 5S)$, pb \\
     \hline
     \multirow{3}{*}{$\Upsilon(\rm 2S)\eta[\gamma\gamma]$} & $10.63$ -- $10.77$ & $3.8$  & $0$ & $2$ & \multirow{6}{*}{$<0.45$} & \multirow{6}{*}{$1.39\pm0.14$} \\
     & $10.83$ -- $10.91$ & $10.1$ & $2.0\pm1.5$ &  $5$ & & \\
     & $10.93$ -- $11.02$ & $7.1$ & $1.0\pm1.0$ & $2$ & & \\
     \cline{1-5}
     
     \multirow{3}{*}{$\Upsilon(\rm 2S)\eta[3\pi]$} & $10.63$ -- $10.77$ & $3.8$  & $1.0\pm1.0$ & $1$ &  &  \\
     & $10.83$ -- $10.91$ & $10.1$ & $17.3\pm4.4$ &  $21$ & & \\
     & $10.93$ -- $11.02$ & $7.1$ & $0$ & $1$ & & \\
     \hline
     
     \multirow{3}{*}{$\Upsilon(\rm 1S)\eta^{\prime}[\pi\pi\eta]$} & $10.63$ -- $10.77$ & $3.8$  & $0$ & $3$ & \multirow{6}{*}{$<0.116$} & \multirow{6}{*}{$<0.021$} \\
     & $10.83$ -- $10.91$ & $10.1$ & $0$ &  $8$ & & \\
     & $10.93$ -- $11.02$ & $7.1$ & $0$ & $8$ & & \\
     \cline{1-5}
     
     \multirow{3}{*}{$\Upsilon(\rm 1S)\eta^{\prime}[\rho\gamma]$} & $10.63$ -- $10.77$ & $3.8$  & $0.8\pm1.2$ & $3$ &  &  \\
     & $10.83$ -- $10.91$ & $10.1$ & $0$ &  $18$ & & \\
     & $10.93$ -- $11.02$ & $7.1$ & $1.3\pm1.8$ & $18$ & & \\
     \hline
      \multirow{3}{*}{$\Upsilon(\rm 1S)\eta[3\pi]$} & $10.63$ -- $10.77$ & $3.8$  & $0$ & $1$ & \multirow{3}{*}{$<0.27$} & \multirow{3}{*}{$0.29\pm0.05$} \\
     & $10.83$ -- $10.91$ & $10.1$ & $0.9\pm1.1$ &  $11$ & & \\
     & $10.93$ -- $11.02$ & $7.1$ & $1.0\pm1.0$ & $3$ & & \\
     
\hline \hline
\end{tabular}
\end{table*}

\subsection{Systematic uncertainties}\label{sect_systematics}

The particle reconstruction efficiency and particle identification are important parameters whose values in simulation could deviate from those in experiment. 
According to independent studies, for example using the $D^{*-}\to \pi^-\Bar{D}^0[K^0_S\pi^+\pi^-]$ decay, the systematic uncertainty due to track reconstruction is $1\%$ for pions and $0.35\%$ for high-momentum muons~\cite{Garmash:2014dhx}.
The photon reconstruction uncertainty is $1.5\%$. 
The muon identification uncertainty is $1\%$ according to analysis of $J/\psi \to \mu^+\mu^-$ \cite{Garmash:2014dhx}.
Therefore, the total systematic uncertainty for the $\mu^+\mu^-\pi^+\pi^-\gamma$ and $\mu^+\mu^-\pi^+\pi^-\gamma\gamma$ final states is $2.7\%$ from charged track reconstruction, $1.5\%$ or $3\%$ from photons reconstruction respectively, and $2\%$ from muon identification.

Another uncertainty can come from the accuracy of the PHOTOS module, which describes final-state radiation. 
To evaluate this uncertainty we simulate the $\Upsilon(\rm 2S)\eta[\gamma\gamma]$ and $\Upsilon(\rm 2S)\eta[3\pi]$ modes without the PHOTOS module.
For both processes the cross section increases by $9\%$ mostly due to absence of radiation by muons, which could account for hundreds of MeV of energy.
Thus, the total influence of PHOTOS on the efficiency is $9\%$ while its own uncertainty is a few percent~\cite{PHOTOS}; therefore, the uncertainty of the detection efficiency appears in the next order and we take $1\%$ as a conservative estimate.

\begin{table*}[htb]
\caption{Systematic uncertainties}
    \label{tab:systematics}
\begin{tabular}
 {@{\hspace{0.5cm}}l@{\hspace{0.2cm}}  @{\hspace{0.2cm}}c@{\hspace{0.5cm}} @{\hspace{0.5cm}}c@{\hspace{0.5cm}} @{\hspace{0.5cm}}c@{\hspace{0.5cm}} @{\hspace{0.5cm}}c@{\hspace{0.5cm}} @{\hspace{0.5cm}}c@{\hspace{0.5cm}}}
\hline \hline
        Uncertainty, $\%$ & $\Upsilon(\rm 2S)\eta[\gamma\gamma]$ & $\Upsilon(\rm 2S)\eta[3\pi]$ &  $\Upsilon(\rm 1S)\eta[3\pi]$ & $\Upsilon(\rm 1S)\eta^{\prime}[\pi\pi\eta]$ & $\Upsilon(\rm 1S)\eta^{\prime}[\rho\gamma]$\\
        \hline
        Track reconstruction & \multicolumn{5}{c}{$2.7$}  \\
        Muon identification & \multicolumn{5}{c}{$2.0$}  \\
        Luminosity $L$ & \multicolumn{5}{c}{$1.4$}  \\
        PHOTOS & \multicolumn{5}{c}{$1.0$}  \\
        
        Radiative correction & $4.3$ & $5.1$ & $5.7$ & $5.7$ & $5.7$ \\
        Photon reconstruction & $3.0$ & $3.0$ & $3.0$ & $3.0$ & $1.5$ \\
        Intermediate branchings & $2.5$ & $8.9$ & $2.4$ & $2.7$ & $2.4$ \\
        Selection criteria & $6.0$ & $6.6$ & $5.6$ & -- & -- \\
        Resolution & $2.1$ & $1.4$ & $1.1$ & -- & -- \\
        Signal lineshape & $1.0$ & $1.4$ & $1.4$ & -- & -- \\
        Background lineshape & $1.5$ & $1.0$ & $1.1$ & -- & -- \\
        Binning & $0.3$ & $2.1$ & $0.8$ & -- & -- \\
        \hline
        Total & $9.6$ & $13.4$ & $9.8$ & $10.0$ & $9.5$\\
\hline \hline
\end{tabular}
\end{table*}

Cross section dependence over energy could differ from the resonance one, leading to systematic uncertainty for the radiative correction.
As alternative dependence we use the sum of the $\Upsilon(\rm 5S)$ Breit-Wigner and the constant contribution, whose amplitude is derived from the upper limit of $0.45$~pb below $\Upsilon(\rm 5S)$ (see Table~\ref{tab:exp_ndet_off}).
The upper limit of $0.45$~pb below $\Upsilon(\rm 5S)$ corresponds to $0.58$~pb after applying the correction for initial-state radiation.
Considering this $0.58$~pb as a constant contribution into Born cross section and using visible cross section  of $1.39$~pb at $\sqrt{s}=10.866$~GeV, one can estimate that the corrected cross section at $\sqrt{s}=10.866$~GeV is $2.10$~pb, implying that a relative amplitude of the constant contribution is $0.58/2.10=0.276$.
Using this cross section dependence we calculate a radiative correction for all modes. 
Its deviation from nominal values ranges from $4.3\%$ to $5.7\%$ and is referred to as a radiative correction uncertainty.

\begin{table}[t]
\caption{Branching fractions used in this work.}
\label{tab:branchings}
\begin{tabular}
 {@{\hspace{0.4cm}}l@{\hspace{0.4cm}}  @{\hspace{0.2cm}}c@{\hspace{0.2cm}} }
\hline \hline
Decay & Branching fraction \cite{PDG}, $\%$ \\
\hline
$\Upsilon(\rm 1S)\to\mu^+\mu^-$ & $2.48\pm0.05$ \\
$\Upsilon(\rm 2S)\to\mu^+\mu^-$ & $1.93\pm0.17$ \\
$\Upsilon(\rm 2S)\to\Upsilon(\rm 1S)\pi^+\pi^-$ & $17.85\pm0.26$ \\
$\eta\to\gamma\gamma$ & $39.41\pm0.2$ \\
$\eta\to\pi^+\pi^-\pi^0$ & $22.92\pm0.28$ \\
$\eta^{\prime}\to\pi^+\pi^-\eta$ & $42.5\pm0.5$ \\
$\eta^{\prime}\to\rho^0\gamma$ & $29.5\pm0.4$ \\
$\pi^0\to\gamma\gamma$ & $98.823\pm0.034$ \\
\hline \hline
\end{tabular}
\end{table}

To estimate the influence of selection criteria we vary three unified 
requirements and check cross section stability.  
The width of the $E_{\rm tot}$ signal range is symmetrically varied by 
$\pm60$~MeV from the nominal value, the lower boundary for the angle 
$\Psi$ is varied from $2$~radian to $2.8$~radian, and the width of the 
$M_{\mu\mu}$ signal range is symmetrically varied by $\pm200$~MeV/$c^2$ from 
the nominal value. 
The maximum cross section deviation from the nominal is taken as a systematic 
uncertainty. 
The total uncertainty due to selection criteria is a quadratic sum of these 
three contributions, and is shown in Table~\ref{tab:systematics}.

One more source of the simulation uncertainty is the deviation between simulated 
and experimental resolutions -- usually experimental distributions are wider 
than those in simulation. 
To evaluate the deviation for systematic uncertainty, we choose events with the $\Upsilon(\rm 1S)$
originating from the $\Upsilon(\rm 2S)\to\Upsilon(\rm 1S)\pi^+\pi^-$ decay using the 
requirement $|M_{\mu\mu\pi\pi}-M_{\mu\mu}-(M_{\Upsilon(\rm 2S)}-M_{\Upsilon(\rm 1S)})|< 18 $~MeV/$c^2$. 
Parameterization of the experimental $M_{\mu\mu}$ distribution with a sum of a 
Gaussian and linear function finds a resolution of $54\pm1.5$~MeV/$c^2$, which is 
larger than the simulated resolution of $50$~MeV/$c^2$ by $8\%$. 
This deviation is common for other distributions; therefore, we vary the resolution of the signal $M_{\eta^{(\prime)}}$ distribution by $\pm10\%$ to calculate the reconstruction efficiency and to fit experimental data. 
The maximum deviation of the cross section from the nominal one is referred to 
as a resolution uncertainty. 
Additionally, we verified that the data parameterization with unfixed resolution 
is consistent with the simulation within statistical uncertainty.

The signal lineshape uncertainty is taken as the maximum difference of the cross section between
data fits with different signal parameterizations.
The nominal lineshape is the sum of the Crystal Ball function and a Gaussian while two tested alternate parameterizations are a Gaussian only and a Crystal Ball only. 
The background lineshape uncertainty is the maximum difference of the cross section 
between data parameterizations taken over different ranges -- in this way not every background event is included in the fit and the background lineshape changes.

The $\eta^{\prime}\to\pi^+\pi^-\eta$ decay was simulated uniformly in phase space, which is not necessarily a correct representation of dynamics of this process.
However, Ref.~\cite{eta-prime} shows that experimental Dalitz distributions are similar to those in the uniformly distributed over phase space model; thus, this source of uncertainty is neglected.

The nominal bin width for the experimental signal distribution is $10$~MeV/$c^2$.
Variation of the width affects the signal yield and leads to systematic uncertainty evaluated by refitting the data with bin widths of $5$, $8$ and $12$~MeV/$c^2$. 

Also, there is luminosity uncertainty of $1.4\%$ and uncertainty of intermediate branching fractions (Table \ref{tab:branchings}).
The total uncertainty is a quadratic sum of all sources. 
For the $\Upsilon(\rm 1S)\eta^{\prime}[\pi\pi\eta]$ and $\Upsilon(\rm 1S)\eta^{\prime}[\rho\gamma]$ modes, some of the uncertainties cannot be evaluated due to zero signal yield. 
Such uncertainties are assumed to be equal to those in the $\Upsilon(\rm 1S)\eta[3\pi]$ mode.

\section{Crosscheck with $\Upsilon(\rm 5S)\to\Upsilon(\rm 2S)[\Upsilon(\rm 1S)\gamma\gamma]\pi^+\pi^-$ } \label{sect3_1}

To validate the analysis procedure we measure the known process 
$e^+e^-\to\Upsilon(\rm 2S)\pi^+\pi^-$, where  $\Upsilon(\rm 2S)$ is reconstructed via 
the decay chain $\Upsilon(\rm 2S)\to\chi_{bJ}(1P)\gamma$, $\rm \chi_{bJ}(1P)\to \Upsilon(\rm 1S)\gamma$, $\Upsilon(\rm 1S)\to\mu^+\mu^-$ and $J=0,1,2$. 
The cross section for this process is measured independently with the $\Upsilon(\rm 2S)\to\mu^+\mu^-$ decay where the statistics of signal events is much higher~\cite{Garmash:2014dhx}.

The analysis procedure is almost the same as for the other modes.
Selection criteria for this process are based on the same set of common variables: an $\Upsilon(\rm 1S)$ meson is reconstructed by the $M_{\mu\mu}$ in the $9.235\text{ GeV}/c^2 < M_{\mu\mu} <9.685\text{ GeV}/c^2$ range, the angle $\Psi>2.6$ radian, 
and the total reconstructed energy $10.75\text{ GeV}<E_{\rm tot}<10.94$~GeV.
In addition, a requirement on the mass recoiling off two charged pions $M^{\rm rec}_{\pi\pi}$ is applied as $|M^{ \rm rec}_{\pi\pi}-M_{\Upsilon(\rm 2S)}|<30$~MeV/$c^2$.
According to MC simulation, the resolution of $M^{\rm rec}_{\pi\pi}$ is $6$~MeV/$c^2$.
This helps to reduce background from the $e^+e^-\to\Upsilon(\rm 1D)\pi^+\pi^-$ process, where $\Upsilon(\rm 1D)\to\chi_{bJ}\gamma$,  $\rm \chi_{bJ}(\rm 1P)\to\Upsilon(\rm 1S)\gamma$. 

The signal distribution for this mode is the largest of two $M_{\mu\mu\gamma}-M_{\mu\mu}$ values.
This variable reconstructs $\rm \chi_{bJ}(1P) \to \Upsilon(\rm 1S)\gamma$, corresponds to 
$M_{\rm \chi_{bJ}}-M_{\Upsilon(\rm 1S)}$, and reduces correlated contributions to the 
resolution from the measurement of muon momentum.
The studied process results in peaks at $399.1$, $432.5$, and $451.9$~MeV/$c^2$ 
for $J=0,~1,~2$, respectively.
Distributions for each $\rm \chi_{bJ}(\rm 1P)$ are fitted to the sum of the Crystal 
Ball function and a Gaussian in the same way as for the other processes.
Reconstruction efficiencies are $\varepsilon_{\rm \chi_{b0}(1P)}=28.12\pm 0.04 \%$, $\rm \varepsilon_{\chi_{b1}(\rm 1P)}=28.68\pm 0.04 \%$, and $\rm \varepsilon_{\chi_{b2}(\rm 1P)}=28.52\pm 0.04 \%$.

The known products of the intermediate branching fractions $\mathcal{B}_{\rm \chi_{bJ}(\rm 1P)}=\mathcal{B}(\Upsilon(\rm 2S)\to\chi_{bJ}(1P)\gamma)\times \mathcal{B}(\chi_{bJ}(1P)\to \Upsilon(\rm 1S)\gamma)$ are $\mathcal{B}_{\rm \chi_{b0}(\rm 1P)}=(7.37\pm1.28)\times 10^{-4}$, $\mathcal{B}_{\rm \chi_{b1}(\rm 1P)}=(242 \pm 19) \times 10^{-4}$, and $\mathcal{B}_{\rm \chi_{b2}(\rm 1P)}=(128\pm 9)\times 10^{-4}$ \cite{PDG}.
The fraction of $\varepsilon_{\rm \chi_{bJ}(\rm 1P)}\times \mathcal{B}_{\rm \chi_{bJ}(1P)}$ is equal to 
$0.029:1:0.527$ ($J=0,~1,~2$) and determines a relative contribution of each 
$\chi_{\rm bJ}(\rm 1P)$ to the total signal lineshape. 
The total MC signal lineshape is the sum of three contributions, with all parameters except normalization factor being fixed for the data analysis.
The total branching fraction weighted with the efficiency is 
$\mathcal{B}_{\Upsilon(\rm 2S)\pi\pi}=\mathcal{B}(\Upsilon(\rm 1S)\to\mu^+\mu^-)\sum \varepsilon_{\rm \chi_{bJ}(1P)}\times \mathcal{B}(\Upsilon(\rm 2S)\to\chi_{bJ}(1P)\gamma)\times \mathcal{B}(\rm \chi_{bJ}(1P)\to \Upsilon(\rm 1S)\gamma) = (2.69 \pm 0.16)\times 10^{-4}$, and is used to calculate the cross section instead of $\varepsilon \mathcal{B}$ (Eq.~\ref{vis_section}).

\begin{figure}[!htp]
  \centering
  \includegraphics[width=8.6cm]{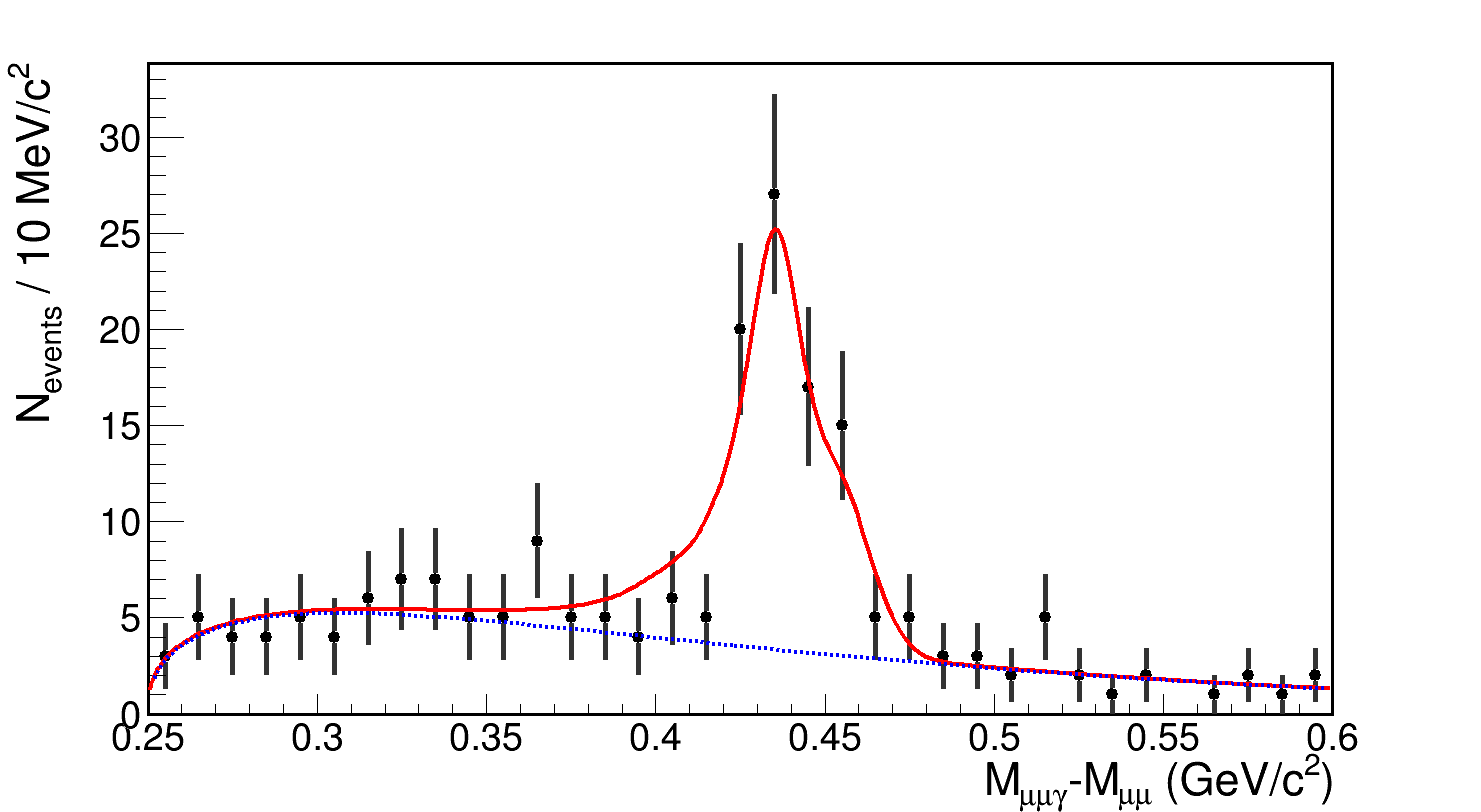} \\  
  \caption{The $M_{\mu\mu\gamma}-M_{\mu\mu}$ distribution for the $e^+e^-\to\Upsilon(\rm 2S)\pi^+\pi^-$ process, where $\Upsilon(\rm 2S)\to\chi_{bJ}(1P)\gamma\to\Upsilon(\rm 1S)\gamma\gamma$. Data are shown as points, the solid red line shows the best fit to the data, and the dashed blue line shows the background contribution.}
    \label{fig:Y2Spcpc_chib}
\end{figure}

Figure~\ref{fig:Y2Spcpc_chib} shows the experimental $M_{\mu\mu\gamma}-M_{\mu\mu}$ distribution.
The signal yield is determined from fitting the $M_{\mu\mu\gamma} - M_{\mu\mu}$ distribution, with the fit function being the sum of the total MC signal lineshape and a background function $(x-p_1)^{p_2}e^{p_3x}$.
We obtain $N_{\rm sig} = 85.32 \pm 11.5$, resulting in the Born cross section $\sigma_{\rm B} (e^+e^-\to\Upsilon(\rm 2S)\pi^+\pi^-) = 3.98 \pm 0.54$~pb (statistical uncertainty only).
This value is consistent with the independent measurement $\sigma_{\rm B} (e^+e^-\to\Upsilon(\rm 2S)\pi^+\pi^-) = 4.07 \pm 0.16 \pm 0.45$ pb~\cite{Garmash:2014dhx} within uncertainty.

\section{Conclusion} \label{sect3_3}
In summary, using the Belle data sample of $118.3$~fb$^{-1}$ obtained at $\sqrt{s}=10.866$~GeV, we report a measurement of the cross section for $e^+e^-\to \Upsilon(\rm 1S,2S)\eta$ processes, and set an upper limit on the cross section of the $e^+e^-\to \Upsilon(\rm 1S)\eta^{\prime}$ process.
The measured Born cross sections, with initial-state radiation being taken into account, are 
the following (Eq. \ref{born crossection}):

$\sigma_{\rm B}^{\eta\to3\pi}(e^+e^- \to \Upsilon(\rm 2S)\eta)=2.08 \pm 0.29 \pm 0.20$~pb,

$\sigma_{\rm B}^{\eta\to2\gamma}(e^+e^- \to \Upsilon(\rm 2S)\eta)=2.07 \pm 0.30 \pm 0.28$~pb,

$\sigma_{\rm B}^{\eta\to3\pi}(e^+e^- \to \Upsilon(\rm 1S)\eta)=0.42 \pm 0.08 \pm 0.04$~pb,

$\sigma_{\rm B}^{\eta^{\prime}\to\pi\pi\eta}(e^+e^- \to \Upsilon(\rm 1S)\eta^{\prime})<0.130$~pb, $CL=90\%$,

$\sigma_{\rm B}^{\eta^{\prime}\to\rho^0\gamma}(e^+e^- \to \Upsilon(\rm 1S)\eta^{\prime})<0.036$~pb, $CL=90\%$.

The weighted averages for the corresponding modes are: 

$\sigma_{\rm B}(e^+e^- \to \Upsilon(\rm 2S)\eta)=2.07 \pm 0.21 \pm 0.19$~pb,

$\sigma_{\rm B}(e^+e^- \to \Upsilon(\rm 1S)\eta)=0.42 \pm 0.08 \pm 0.04$~pb,

$\sigma_{\rm B}(e^+e^- \to \Upsilon(\rm 1S)\eta^{\prime})<0.035$~pb, $CL=90\%$.

Significance being $10.2\sigma$ and $16.5\sigma$ for the $e^+e^- \to \Upsilon(\rm 1S)\eta$ and $e^+e^- \to \Upsilon(\rm 2S)\eta$ processes respectively, we claim the first observation of these processes.
For $e^+e^- \to \Upsilon(\rm 2S)\eta$, the measured cross section deviates from the inclusive measurement \cite{Tamponi:2018cuf} at $\sim2.3\sigma$ considering both statistical and uncorrelated systematic uncertainties.  
For $e^+e^- \to \Upsilon(\rm 1S)\eta$, inclusive upper limit does not contradict our measurement. 

Under the assumption that processes proceed only through the $\Upsilon(\rm 5S)$ meson, we calculate branching fractions with the formula $\mathcal{B}(\Upsilon(\rm 5S)\to X) = \sigma_{\rm vis}(e^+e^-\to X) / \sigma(e^+e^-\to \Upsilon(\rm 5S))$, where $\sigma_{\rm vis}(e^+e^-\to \Upsilon(\rm 5S)) = 0.340\pm0.016$~nb \cite{Esen:2012yz}:

$\mathcal{B}(\Upsilon(\rm 5S)\to \Upsilon(\rm 1S)\eta ) = (0.85 \pm 0.15 \pm 0.08)\times 10^{-3}$,

$\mathcal{B}(\Upsilon(\rm 5S)\to \Upsilon(\rm 2S)\eta ) = (4.13 \pm 0.41 \pm 0.37)\times 10^{-3}$,

$\mathcal{B}(\Upsilon(\rm 5S)\to \Upsilon(\rm 1S)\eta^{\prime} ) < 6.9 \times 10^{-5}$, $CL=90\%$.

Using $\sigma(e^+e^- \to \Upsilon(\rm 1S)\pi^+\pi^-)=2.27 \pm 0.12 \pm 0.14$~pb, $\sigma(e^+e^- \to \Upsilon(\rm 2S)\pi^+\pi^-)=4.07 \pm 0.16 \pm 0.45$~pb \cite{Garmash:2014dhx} and the obtained Born cross section, we also calculate the width ratios between $\eta$- and dipion-transitions to be 
\begin{equation}
     \frac{\Gamma(\Upsilon(\rm 5S)\to\Upsilon(\rm 1S)\eta)}{\Gamma(\Upsilon(\rm 5S)\to\Upsilon(\rm 1S)\pi^+\pi^-)}= 0.19\pm 0.04\pm 0.01
\end{equation}
and
\begin{equation}
    \frac{\Gamma(\Upsilon(\rm 5S)\to\Upsilon(\rm 2S)\eta)}{\Gamma(\Upsilon(\rm 5S)\to\Upsilon(\rm 2S)\pi^+\pi^-)}= 0.51\pm 0.06 \pm 0.04 ,
\end{equation}
where correlated systematic uncertainties are canceled out.
These values are noticeably larger than the predicted values of $\sim0.03$ for $\Upsilon(\rm 2S)$ and $\sim0.005$ for $\Upsilon(\rm 1S)$, calculated in the QCDME regime, see Ref.~\cite{Voloshin:1980zf}, and are comparable to the $\frac{\Upsilon(\rm 4S)\to\Upsilon(\rm 1S)\eta}{\Upsilon(\rm 4S)\to\Upsilon(\rm 1S)\pi^+\pi^-} =2.41\pm0.40\pm0.12$~\cite{Aubert:2008az}, measured in a regime where QCDME is no longer valid.
Similarly, our measured upper limit on the ratio between the $\eta^{\prime}$ and $\eta$ transitions is 
\begin{equation}
    \frac{\Gamma(\Upsilon(\rm 5S)\to\Upsilon(\rm 1S)\eta^{\prime})}{\Gamma(\Upsilon(\rm 5S)\to\Upsilon(\rm 1S)\eta)} < 0.09 ~(CL=90\%),
\end{equation}
which is significantly smaller than the value of $\sim 12$ predicted by the naive QCDME model \cite{Voloshin:2011hw}.

As shown in Refs. \cite{Voloshin:2012dk, Voloshin:2011hw}, one of the possible 
solutions is existence of a light-flavor admixture to the $b\Bar{b}$ state.
Such a structure of the $\Upsilon(\rm 5S)$ resonance could increase the cross section of $e^+e^-\to \Upsilon(\rm 1S,2S)\eta$ and $e^+e^-\to \Upsilon(\rm 1S)\eta^{\prime}$ processes and lead to dominance of the $e^+e^-\to \Upsilon(\rm 1S,2S)\eta$ process over $e^+e^-\to \Upsilon(\rm 1S)\eta^{\prime}$ \cite{Voloshin:2012dk}:
\begin{equation}
    \frac{\Gamma(\Upsilon(\rm 5S)\to\Upsilon(\rm 1S)\eta^{\prime})}{\Gamma(\Upsilon(\rm 5S)\to\Upsilon(\rm 1S)\eta)} \approx \frac{p^3_{\eta^{\prime}}}{2p^3_{\eta}} = 0.25, 
\end{equation}
that is much higher than the obtained limit.
Such suppression also has been observed in Ref. \cite{Guido:2018ywg}, where $\frac{\Gamma(\Upsilon(\rm 4S)\to\Upsilon(\rm 1S)\eta^{\prime})}{\Gamma(\Upsilon(\rm 4S)\to\Upsilon(\rm 1S)\eta)}$ is reported to be $0.20 \pm 0.06$, in agreement with the expected value in the case of an admixture of a state containing light quarks.


\section{Acknowledgement}

We thank the KEKB group for the excellent operation of the
accelerator; the KEK cryogenics group for the efficient
operation of the solenoid; and the KEK computer group, and the Pacific Northwest National
Laboratory (PNNL) Environmental Molecular Sciences Laboratory (EMSL)
computing group for strong computing support; and the National
Institute of Informatics, and Science Information NETwork 5 (SINET5) for
valuable network support.  We acknowledge support from
the Ministry of Education, Culture, Sports, Science, and
Technology (MEXT) of Japan, the Japan Society for the 
Promotion of Science (JSPS), and the Tau-Lepton Physics 
Research Center of Nagoya University; 
the Australian Research Council including grants
DP180102629, 
DP170102389, 
DP170102204, 
DP150103061, 
FT130100303; 
Austrian Federal Ministry of Education, Science and Research (FWF) and
FWF Austrian Science Fund No.~P~31361-N36;
the National Natural Science Foundation of China under Contracts
No.~11435013,  
No.~11475187,  
No.~11521505,  
No.~11575017,  
No.~11675166,  
No.~11705209;  
Key Research Program of Frontier Sciences, Chinese Academy of Sciences (CAS), Grant No.~QYZDJ-SSW-SLH011; 
the  CAS Center for Excellence in Particle Physics (CCEPP); 
the Shanghai Science and Technology Committee (STCSM) under Grant No.~19ZR1403000; 
the Ministry of Education, Youth and Sports of the Czech
Republic under Contract No.~LTT17020;
Horizon 2020 ERC Advanced Grant No.~884719 and ERC Starting Grant No.~947006 ``InterLeptons'' (European Union);
the Carl Zeiss Foundation, the Deutsche Forschungsgemeinschaft, the
Excellence Cluster Universe, and the VolkswagenStiftung;
the Department of Atomic Energy (Project Identification No. RTI 4002) and the Department of Science and Technology of India; 
the Istituto Nazionale di Fisica Nucleare of Italy; 
National Research Foundation (NRF) of Korea Grant
Nos.~2016R1\-D1A1B\-01010135, 2016R1\-D1A1B\-02012900, 2018R1\-A2B\-3003643,
2018R1\-A6A1A\-06024970, 2019K1\-A3A7A\-09033840,
2019R1\-I1A3A\-01058933, 2021R1\-A6A1A\-03043957,
2021R1\-F1A\-1060423, 2021R1\-F1A\-1064008;
Radiation Science Research Institute, Foreign Large-size Research Facility Application Supporting project, the Global Science Experimental Data Hub Center of the Korea Institute of Science and Technology Information and KREONET/GLORIAD;
the Polish Ministry of Science and Higher Education and 
the National Science Center;
the Ministry of Science and Higher Education of the Russian Federation, Agreement 14.W03.31.0026, 
and the HSE University Basic Research Program, Moscow; 
University of Tabuk research grants
S-1440-0321, S-0256-1438, and S-0280-1439 (Saudi Arabia);
the Slovenian Research Agency Grant Nos. J1-9124 and P1-0135;
Ikerbasque, Basque Foundation for Science, Spain;
the Swiss National Science Foundation; 
the Ministry of Education and the Ministry of Science and Technology of Taiwan;
and the United States Department of Energy and the National Science Foundation.

\bibliography{ekoval}

\end{document}